\documentclass[11pt]{article}
\usepackage{amssymb,srcltx,epsfig,bm,color,rotating}
\usepackage{breakcites, comment}
\usepackage{multirow,url}
\usepackage[centertags]{amsmath}
\usepackage{subfigure}
\usepackage{breakcites}

\usepackage{array}
\usepackage[]{natbib}
\usepackage{amssymb,srcltx,epsfig,bm,color,rotating,amsmath}
\usepackage{booktabs, setspace} 
\usepackage{eps2pdf}
\usepackage[english]{babel}

\usepackage{booktabs, multirow, tabularx}

\newcommand{\makecell}[2][@{}c@{}]{\begin{tabular}{#1}#2\end{tabular}}

\definecolor{DarkGreen}{rgb}{0.5,0.8,0.6}   
\definecolor{RGBblack}{rgb}{0.0,0.0,0.0}    

\def\stackover#1#2{\mathrel{\mathop{#2}\limits^{#1}}}
\newcommand{\iid}{\stackover{\mbox{\footnotesize i.i.d.}}{\sim}}

\newcommand{\is}{\itemsep=0pt}
\newcommand{\bd}[1]{\begin{description}[#1]\is}
\newcommand{\ed}{\end{description}}
\newcommand{\bi}{\begin{itemize}\is}
  \newcommand{\ei}{\end{itemize}}
\newcommand{\be}{\begin{enumerate}\is}
  \newcommand{\ee}{\end{enumerate}}
  \newcommand{\beq}{\begin{eqnarray}\is}
  \newcommand{\eeq}{\end{eqnarray}}
\makeatletter
\newcommand*{\rom}[1]{\expandafter\@slowromancap\romannumeral #1@}

\newcommand{\eb}{\bm e}

\newcommand{\bz}{\bm z}
\newcommand{\br}{\bm r}

\newcommand{\bx}{\bm x}
\newcommand{\by}{\bm y}

\newcommand{\bgamma}{\bm \gamma}

\newcommand{\bmu}{\bm \mu}
\newcommand{\bbeta}{\bm \beta}

\usepackage{fullpage}  
\usepackage{caption}

\usepackage{graphicx}

\begin{document}
\doublespacing

\title{BAREB: A Bayesian repulsive biclustering model for periodontal data}

\author{Yuliang Li$^{1}$, Dipankar Bandyopadhyay$^{2}$, Fangzheng Xie$^{1}$, Yanxun Xu$^{1,*}$\\
$^{1}$Department of Applied Mathematics and Statistics, Johns Hopkins University, U.S.A. \\
$^{2}$Department of Biostatistics, Virginia Commonwealth University\\
$^{*}$ to whom correspondence should be addressed
}

\date{}
\maketitle


\begin{abstract}
Preventing periodontal diseases (PD) and maintaining the structure and function of teeth are important goals for personal oral care. To understand the heterogeneity in patients with diverse PD patterns, we develop BAREB, a Bayesian repulsive biclustering method that can simultaneously cluster the PD patients and their tooth sites after taking the patient- and site-level covariates into consideration. BAREB uses the determinantal point process (DPP) prior to induce diversity among different biclusters to facilitate parsimony and interpretability. Since PD progression is hypothesized to be spatially-referenced, BAREB factors in the spatial dependence among tooth sites.
In addition, since PD is the leading cause for tooth loss, the missing data mechanism is non-ignorable. Such nonrandom missingness is incorporated into BAREB.  For the posterior inference, we design an efficient reversible jump Markov chain Monte Carlo sampler. Simulation studies show that BAREB is able to accurately estimate the  biclusters, and compares favorably to alternatives. For real world application, we apply BAREB to a dataset from a clinical PD study, and obtain desirable and interpretable results. A major contribution of this paper is the \texttt{Rcpp} implementation of BAREB, available at \url{https://github.com/YanxunXu/BAREB}.

\end{abstract}

\section{Introduction}
\label{sec:intro}
Periodontal disease (PD), a chronic widespread inflammatory disease, can damage the soft tissues and bones that support the teeth, leading to loosening and eventual loss of  teeth. In addition to impacting the quality of life (\citealp{ferreira2017impact}), PD has been linked to a number of systemic diseases, such as heart diseases (\citealp{bahekar2007prevalence}) and diabetes (\citealp{fernandes2009periodontal}). Therefore, preventing PD, and maintaining the structure and function of teeth are important goals of personal oral care. The most popular clinical biomarker quantifying the progression of PD is the clinical attachment level (CAL), defined as the depth (in mm) from the cementoenamel junction to the base of a tooth. In clinical studies, the CAL is measured at six pre-specified tooth-sites (excluding the four third molars) using a periodontal probe, which leads to 168 measurements for a full mouth with no missing teeth (for illustration, see Figure F1 in the Supplementary Materials).

PD data are complex and multi-level (patient-, tooth-, site-level), and traditional summary-based statistical approaches, such as mean (\citealp{pilgram2002relationships}), sum scores, or maximum site-level values when applied to patient-level evaluation lead to imminent loss of information (\citealp{cho2015analysis, nomura2017site}). To mitigate this, and other salient features of PD data such as non-random missingness, spatial association, and non-Gaussianity of responses, \citet{reich2010latent} and \citet{reich2013nonparametric} proposed Bayesian inference under the desired mixed-effects modeling framework. However, in estimating covariate effects on the CAL response, the authors assumed all patients share the same coefficient. This assumption is questionable, as the rate of PD progression can be very different among patients, with possible \textit{clustering} of patients according to PD incidence.

The motivation for this work comes from a clinical study of oral PD assessment among Gullah-speaking African-Americans (henceforth, GAAD study) residing in the coastal South Carolina sea-islands (\citealp{fernandes2009periodontal}). To evaluate the heterogeneity of PD incidence among patients, available one-dimensional clustering methods focus on grouping either the patients, or their tooth-sites, separately (\citealp{bandyopadhyay2016non}), according to disease status.  While useful, such clustering techniques cannot identify co-localized tooth sites that are important in inferring patient-level clustering. Furthermore, clustering of tooth sites should depend on which subgroup of patients we focus upon, given that different subgroups may partition the tooth sites in different ways, indicating different PD patterns. Therefore, it is desirable to learn whether there exist subgroups of patients, such that within each subgroup, the PD incidence of some tooth sites are different from others, and also different from patients in other subgroups. However, this inferential framework is further complicated in presence of missing data. In the GAAD dataset, a considerable proportion of patients (around 95\%) have missing teeth, with an average of 32\% teeth missing for a patient. This missingness is often assumed \textit{non-ignorable} in PD studies (\citealp{reich2010latent}), since PD is a major cause for tooth-loss, and estimating the counterfactual CAL values the missing tooth-sites (from a missing tooth) would have had if the tooth was not missing can facilitate subgroup identification. This is important, in conjunction to spatial associations observed in PD progression studies (\citealp{reich2010latent}), i.e., proximally located tooth/tooth-sites may exhibit similar PD patterns compared to the distally located ones, because a missing/failed tooth is predictive of higher PD status, and often expected to be surrounded by teeth with high CAL (\citealp{schnell2015marginal}). We aim to fill these gaps in the existing PD literature by developing a probabilistic \textit{biclustering}, or two-dimensional clustering (\citealp{cheng2000biclustering}) method.

The current literature on biclustering is considerably rich (\citealp{getz2000coupled, gu2008bayesian, li2009qubic, lee2013nonparametric}), with applications to genomics and other fields. However, all these biclustering methods focus on the \textit{mean} in each bicluster, and fails to incorporate important covariate information (such as age, gender), and aforementioned data characteristics typical to PD. Also, from a Bayesian standpoint, utilizing independent priors on the bicluster-specific parameters continues to remain popular due to their computational convenience and flexibility (\citealp{gu2008bayesian, lee2013nonparametric, xu2013nonparametric}). However, such an approach could cause over-fitting issues and redundant biclusters, leading to inferences that are hard to interpret (\citealp{xie2019bayesian}). For example, the NoB-LCP biclustering method of \citet{xu2013nonparametric} used the Dirichlet process (DP) priors, where the atoms are independent and identically distributed (i.i.d.) from a base distribution, to infer clustering of histone modifications and genomic locations. Due to the properties of DP, NoB-LCP inferred a large number of small clusters with very few genomic locations, leading to unnecessarily complex models and poor interpretability.

To this end, we develop BAREB, a \textbf{BA}yesian \textbf{RE}pulsive \textbf{B}iclustering model to study the heterogeneity in patients with diverse PD patterns. Our contributions are three fold. First, under a matrix formulation of our CAL responses (with rows as patients and columns as tooth-sites), the proposed method produces simultaneous clustering of the study patients and their tooth sites, taking into account the spatial association among tooth sites and non-random missingness patterns, and provides model-based posterior probabilities for these random partitions. These biclusters are defined via consistent associations between CAL values and covariates among a subset of tooth sites for a subgroup of patients. In other words, our proposed model will cluster any two patients together, if they give rise to the same partition of tooth sites. Second, to address the issues with independent priors, we make use of a repulsive prior -- the determinantal point process, or DPP (\citealp{macchi1975coincidence}) on the random partitions, which encourages diversity in PD patterns among different biclusters. Bayesian inference using the DPP priors have proved to be extremely effective in facilitating parsimony and interpretability in a variety of mixture and latent feature allocation models with biomedical applications \citet{affandi2013approximate, xu2016bayesian} to infer clinically meaningful subpopulations. Third, BAREB introduces a (latent) shared-parameter framework  to deal with the non-ignorable missingness, with the resulting marginal mixture density accommodating the non-Gaussianity of CAL responses. In addition, integrated \texttt{R} and \texttt{C++} codes for implementing BAREB in other application domains are available via \texttt{GitHub}, to be eventually submitted as a \texttt{R} package in \texttt{CRAN}.

The rest of the paper proceeds as follows. In Section \ref{sec:model}, we present the statistical formulation for BAREB. Section \ref{sec:bayes} develops the Bayesian inferential framework, with the associated choice of priors, joint likelihood, posteriors, and model comparison measures. In Section \ref{sec:sim}, we evaluate the finite sample performance of BAREB, and the advantages of using the repulsive DPP over plausible alternatives via a simulation study. Application of our method to the motivating GAAD data is presented in Section \ref{sec:apply}. Finally, we conclude, with a discussion in Section \ref{sec:end}.

\section{Statistical Model} 
\label{sec:model}

We use an $N \times J$ matrix $Y=[y_{ij}]$ to represent the observed CAL values, with $y_{ij}$ denoting the CAL for patient $i$ at tooth site $j$, where $i=1,\ldots, N; j=1,\ldots, J$. From the motivating GAAD study, we consider patients with at least one tooth present, and with complete set of covariates. Missing CAL values are denoted as $y_{ij}=\mathrm{NA}$.
Note, a tooth-site is missing, if and only if the corresponding tooth is missing. Hence, we consider the missingness indicator $\delta_i(t)$ at the tooth-level, i.e., $\delta_i(t)=1$ if tooth $t$ of patient $i$ is missing, otherwise $\delta_i(t)=0$, $i=1, \dots, N$; $t=1, \dots, T$.

BAREB clusters any two patients together if they have the same partition of tooth sites after accounting for patient-level covariates. Since the clustering of tooth sites are nested within clusters of patients, we start the model construction with a random partition of patients $\{1, \ldots, N\}$ by denoting the vector $\eb=(e_{1}, \ldots, e_{N})$ as the patient cluster membership indicator. Denote $S$ to be the number of patient clusters, where $e_i=s$ indicates patient $i$ belongs to patient cluster $s$, $s=1, \ldots, S$. We propose a categorical distribution prior for $\eb$, such that
\begin{equation}
e_i\iid \mathrm{Categorical}(\bm{w}), \ \ i=1, \dots, N,
\label{eq:priore}
\end{equation}
\noindent where $\bm{w}=(w_1, \ldots, w_S)$ with $\sum_{s=1}^Sw_s=1$. We assume a Dirichlet distribution prior on $\bm{w}$, such that $\bm{w}\sim \text{Dirichlet}(\bm{\alpha})$, where $ \bm{\alpha}=(\alpha_1, \ldots, \alpha_S)$.

Next, we consider clustering of tooth sites for each of the $S$ patient clusters. Recall, the partition of tooth sites is nested within patient clusters, i.e., site clusters can be different for different patient clusters. Let $D_{s}$ be the number of tooth site clusters for the $s$-th patient cluster. Define $\br_{s}=(r_{s1}, \ldots, r_{sJ})$ the vector of clustering labels $r_{sj} \in \{1, \dots, D_s \}$ that describe the partition of tooth sites corresponding to the $s$-th patient cluster, where $r_{sj}=d$ denotes that tooth site $j$ is assigned to site cluster $d$ in patient cluster $s$. Letting $\br=(\br_1, \ldots, \br_S)$, we assume independent categorical priors for each $\br_{s}$, given by
\begin{equation}
p(\br\mid \eb)=\prod_{s=1}^Sp(\br_s) \ \ \ \mathrm{and} \ \ \ \ r_{sj}\iid\mathrm{Categorical}(\bm{\phi}_s), \ j=1, \ldots, J,
\label{eq:priorr}
\end{equation}
\noindent where $\bm{\phi_{s}}\sim \text{Dirichlet}(\bm{\alpha^{\phi}_{s}})$, with $\bm{\alpha^{\phi}_{s}} = (\alpha^{\phi}_{s1}, \ldots, \alpha^{\phi}_{sD_s})$.
The prior probability models on the biclustering in \eqref{eq:priore} and \eqref{eq:priorr} can be characterized as a partition of patients and a nested partition of tooth sites, nested within each cluster of patients. These biclusters will provide valuable information on the periodontal decay patterns of teeth and heterogeneity among PD patients for developing subsequent prevention and treatment strategies.

Given $\eb$ and $\br$, we construct a sampling model for the CAL as:
\begin{eqnarray}
y_{ij}= \bx_{i}\bbeta_{i}+\bz_{j}\bgamma_{ij} + \nu_{ij} + \epsilon_{ij}. 
\label{eq:mainmodel}
\end{eqnarray}
Here, $\bx_i$ is the vector of patient-level covariates (e.g., age) for patient $i$ with the corresponding regression parameter $\bbeta_{i}$; $\bz_j$ is the vector of tooth site-level covariates (e.g., jaw indicator) including an intercept term with the corresponding regression parameter $\bgamma_{ij}$;  $\nu_{ij}$ models the spatial dependence among tooth sites; and $\epsilon_{ij}$ are independent errors: $\epsilon_{ij}\iid N(0,\sigma^{2})$.

\begin{figure}[hbt!]
\begin{center}
\includegraphics[scale=0.6]{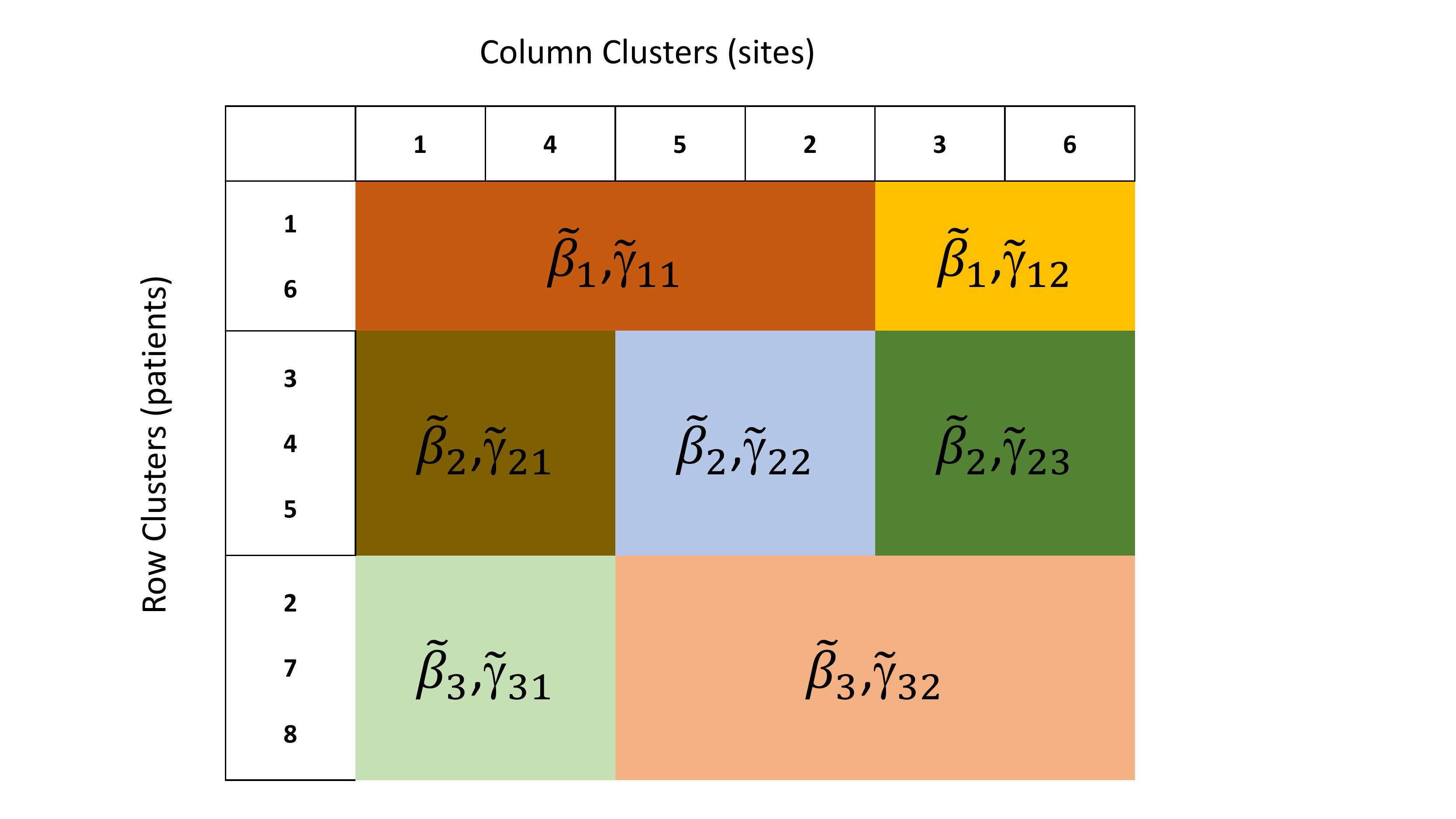}
\caption{An illustration of the proposed BAREB model with 8 patients and 6 tooth sites. Here, we assume 3 patient clusters.}
\label{fig:bicluster}
\end{center}
\end{figure}

The prior probability models for $\bbeta_{i}$ and $\bgamma_{ij}$ make use of the biclustering. We define $p(\bbeta_{i}, \bgamma_{ij}\mid \eb, \br$) as follows. If $e_i=s$ and $r_{sj}=d$, we assume all sites in the same site cluster $d$ in patient cluster $s$ share the parameter $\widetilde{\bgamma}_{sd}$, and all patients in the same patient cluster $s$ share the parameter $\widetilde{\bbeta}_{s}$, i.e., $\bbeta_i=\widetilde{\bbeta}_{s}$ for all $i$ with $e_i=s$ and $\bgamma_{ij}=\widetilde{\bgamma}_{sd}$ for all $i$ and $j$, with $e_i=s$ and $r_{sj}=d$.
Figure~\ref{fig:bicluster} presents a graphical illustration of the proposed BAREB model with 8 patients and 6 tooth sites. Here, we assume three patient clusters, with cluster \# 1 having two site clusters, cluster \# 2 having three site clusters, and cluster \# 3 having two site clusters. Varying colors indicate the different biclusters with the corresponding parameters. Within each patient cluster, all patients share the same $\widetilde{\bbeta}_s$, but different $\widetilde{\bgamma}_{sd}$'s across different site clusters, with $d=1, \ldots, D_s$; $s=1, \ldots, S$. We will discuss the priors for $\widetilde{\bbeta}_s$ and $\widetilde{\bgamma}_{sd}$ in Section \ref{sec:bayes}.

To account for non-randomly missing data, we introduce a latent variable $g_i(t)$, such that $\delta_i(t)=I(g_i(t)>0)$, $t=1, \dots, T$. We model $g_i(t)$ as
\begin{eqnarray}
g_{i}(t)\sim N(\mu_{i}^{*}(t),1) \ \ \ \mathrm{and} \ \   \ \mu_{i}^{*}(t) = c_0 + c_1 R_t\bmu_i,
\label{eq:missing}
\end{eqnarray}
\noindent where $\bmu_i = (\mu_{i1}, \ldots, \mu_{iJ})$ with each element $\mu_{ij}=\bx_{i}\bbeta_{i}+\bz_{j}\bgamma_{ij} + \nu_{ij}$, $R_t$ is a $J$-dimensional vector with $R_t(j)=\frac{1}{6}$ if site $j$ is on tooth $t$, and $0$ otherwise, and $\bm c = (c_0, c_1)$ is the unknown (estimable) parameter that controls the relationship between the CAL response and the probability of missing tooth. Under this shared-parameter joint modeling framework, the mean $\bmu_i$ is shared between the two regression models, with the mean of $g_{i}(t)$ specified as the average CAL value corresponding to the six locations of tooth $t$ for patient $i$. If $c_1>0$, higher CAL values will more likely result into increased probability of missing tooth, and vice versa. From \eqref{eq:missing}, we can easily derive $p(\delta_{i}(t)=1)=\Phi(\mu_{i}^{*}(t))$ after integrating out $g_i(t)$, where $\Phi(\cdot)$ is the cumulative distribution function of the standard normal distribution.

\section{Bayesian Inference}
 \label{sec:bayes}
\subsection{Priors and joint likelihood}
In this subsection, we discuss the priors on the linear coefficients $\widetilde{\bbeta}_s$'s and $\widetilde{\bgamma}_{sd}$'s, and the spatial term $\nu_{ij}$.
In practice, independent priors on biclustering-specific parameters $\widetilde{\bbeta}_s$'s and $\widetilde{\bgamma}_{sd}$'s are preferred due to their computational tractability. However, such an approach could cause redundant clusters, resulting in inferences that are hard to interpret as biologically/clinically meaningful subpopulations (\citealp{xu2013nonparametric,lee2013nonparametric}). \citet{xu2016bayesian} proposed to use the DPP as a prior to induce repulsiveness among component-specific parameters in the context of Gaussian mixture models, and showed that the DPP prior yields more parsimonious and interpretable inference compared to independent priors. Now, we extend the use of the DPP in our biclustering setup to encourage repulsive coefficients $\widetilde{\bbeta}_s$'s and $\widetilde{\bgamma}_{sd}$'s, thereby inducing diverse PD patterns among different biclusters.

Let $C^{\beta}$ denote an $S\times S$ positive semidefinite matrix constructed through a covariance function  $C^{\beta}_{ss'}=C^{\beta}(\bbeta_s, \bbeta_{s'})$. We slightly modify the DPP prior used in \citet{xu2016bayesian} by defining a prior on $(\widetilde{\bbeta}_1, \ldots, \widetilde{\bbeta}_S)$, with respect to the $S$-dimensional Lebesgue measure on $\mathbb{R}^S$ as
\begin{eqnarray}
p(\widetilde{\bbeta}_1, \dots, \widetilde{\bbeta}_S\mid S) = \frac{1}{Z_S}\mathrm{det}[C^{\beta}](\widetilde{\bbeta}_1, \dots, \widetilde{\bbeta}_S)
\label{eq:rep1}
\end{eqnarray}
\noindent where $Z_S$ is the normalizing constant, and $\mathrm{det}[C^{\beta}](\widetilde{\bbeta}_1, \ldots, \widetilde{\bbeta}_S)$ is the determinant of the matrix $[C^{\beta}_{ss'}]_{S\times S}$. Geometrically, the determinant can be interpreted as the volume of a parallelotope spanned by the column vectors of $C^{\beta}$. Therefore, the prior in  \eqref{eq:rep1} defines a repulsive point process, since equal or similar column vectors span smaller volume than very diverse ones. Specifically, $p(\widetilde{\bbeta}_1, \ldots, \widetilde{\bbeta}_S\mid S)=0$, whenever $\widetilde{\bbeta}_s=\widetilde{\bbeta}_{s'}$ for some $s\neq s'$. In other words, it assigns vanishing values to the density at the point configurations that have replicate(s) within themselves. In this paper, we use $C^{\beta}(\widetilde{\bbeta}_s,\widetilde{\bbeta}_{s'}) = \exp\left \{-\frac{||\widetilde{\bbeta}_s-\widetilde{\bbeta}_{s'}||^2_2}{\theta_{\beta}^2} \right \}$, where $\theta_{\beta}$ is an unknown parameter that controls how repulsive the prior is. If desired, alternative covariance functions can be implemented here without complicating the model. Similarly, we define the repulsive prior on $\widetilde{\bgamma}_{sd}$, $s=1, \ldots, S$, as follows:
\begin{eqnarray}
p(\widetilde{\bgamma}_{s1}, \dots, \widetilde{\bgamma}_{sD_s}\mid D_s) = \frac{1}{Z_{\gamma_s}}\mathrm{det}[C^{\gamma_{s}}](\widetilde{\bgamma}_{s1}, \dots, \widetilde{\bgamma}_{sD_s}),
\label{eq:rep2}
\end{eqnarray}
where $C^{\gamma_s}(\widetilde{\bgamma}_{sd}, \widetilde{\bgamma}_{sd'})=\exp\left \{ -\frac{||\widetilde{\bgamma}_{sd}-\widetilde{\bgamma}_{sd'}||^2_2}{\theta_{\gamma_s}^2} \right \}$. We further assume the priors
$p(D_s)\propto Z_{\gamma_s}/D_s!$, $p(\theta_{\beta})=N(0, \sigma^2_{\theta_{\beta}})$, $p(\theta_{\gamma_s})=N(0, \sigma^2_{\theta_{\gamma}})$, $s=1, \dots, S$.

We assume the priors  $\sigma_{sp}^2\sim \text{IG}(a_{sp}, b_{sp})$ and $\rho\sim \text{Unif}(a_{\rho}, b_{\rho})$, where IG$(\cdot, \cdot)$ and Unif$(\cdot, \cdot)$ denote the Inverse Gamma and Uniform densities, respectively.  For the variance of the error $\epsilon_{ij}$, we assume a conjugate prior $p(\sigma^2)\sim \mathrm{IG}(a_{\sigma^2}, b_{\sigma^2})$. The elements of the parameter vector $\bm c$ determining non-random missingness are assigned conjugate normal priors, i.e., $p(c_{i}) = N(0, \sigma^2_{c_{0}}), i = 0,1$. Note, we do not assume a prior on the number of patient clusters $S$, as it complicates posterior computation if we allow both $S$ and $D_s$, $s=1,\dots, S$ to be random. We will discuss how to choose $S$ in Section \ref{s:post}.

Finally, we complete the model construction by assigning a prior to the spatial term $\nu_{ij}$. Let $\bm\nu_i = (\nu_{i1},\dots,\nu_{iJ})'$. For each patient $i$, we assume
\begin{eqnarray}
\bm\nu_i \iid \text{MVN}(0,\bm\Sigma),
\label{eq:rep3}
\end{eqnarray}
where MVN($\cdot, \cdot$) denotes a multivariate normal density. Following \citet{besag1974spatial}, we model the spatial effect by considering a conditional autoregressive (CAR) prior: $\bm\Sigma = \sigma_{sp}^2 \bm G(\rho)^{-1}$, where $\bm G(\rho) = \bm B- \rho \bm W$. Here $\bm B$ is a diagonal matrix with the $j$th diagonal entry being the number of neighbors at $j$th site and $\bm W$ denotes the adjacency matrix for 168 tooth sites in the mouth structure. To construct the adjacency matrix, we consider the adjacent sites on the same tooth and sites that share a gap between teeth as ‘neighbors’. The adjacency structure considered in both data analysis and simulation studies is presented in Figure F2, Supplementary Materials (see Type I \& II neighbors).

In summary, the joint model of BAREB factors as
\begin{eqnarray*}
&&\underbrace{p(Y\mid \{\widetilde{\bbeta}_s\}_{s=1}^S, \{\widetilde{\bgamma}_{sd}\}_{d=1,s=1}^{D_s, \ \ S}, \{\bm\nu_i\}_{i=1}^N, \eb, \br, \sigma^{2})}_{\eqref{eq:mainmodel}} \underbrace{p(\eb\mid S, \bm{w})}_{\eqref{eq:priore}} \underbrace{p(\br \mid \eb, S, \{\bm{\phi}_s\}_{s=1}^S)}_{\eqref{eq:priorr}}
\underbrace{p(\delta_i(t)\mid c, \mu_i^{*}(t))}_{\eqref{eq:missing}}  \nonumber\\
&& \times \underbrace{p(\{\widetilde{\bbeta}_s\}_{s=1}^S)}_{\eqref{eq:rep1}}
 p(\sigma^2)p(\bm c) p(\theta_{\beta}) p(S)\prod_{s=1}^S \underbrace{p(\{\widetilde{\bgamma}_{sd}\}_{d=1}^{D_s})}_{\eqref{eq:rep2}} p(D_s)p(\theta_{\gamma_s})
 \underbrace{p(\{\bm\nu_i\}_{i=1}^N\mid \sigma_{sp}^2, \rho)}_{\eqref{eq:rep3}}
 p(\sigma_{sp}^2)p(\rho) . \nonumber\\
\end{eqnarray*}

\subsection{Posterior inference} \label{s:post}
We carry out Markov chain Monte Carlo (MCMC) simulations for posterior inference.  One challenging step is to update the number of site clusters $D_s$ nested within each patient cluster. Following \citet{xu2016bayesian}, we design a reversible jump MCMC (RJMCMC) sampler (\citealp{green1995reversible}) that allows random $D_s$ within the DPP prior for  $(\widetilde{\bgamma}_{s1}, \dots, \widetilde{\bgamma}_{sD_s})$ using the moment-matching principle (\citealp{zhang2004learning}) in a multivariate setting. Details of the sampling procedure are described in the Supplementary Materials.

To determine the patient cluster cardinality $S$, we use a model selection procedure based on the Watanabe-Akaike information criterion, WAIC (\citealp{vehtari2017practical}), instead of designing a RJMCMC sampler for computational efficiency. Compared to other popular model selection methods such as AIC, BIC, and DIC (\citealp{spiegelhalter2002bayesian}), the WAIC based on point-wise predictive density is fast, computationally convenient, and fully Bayesian using the full posterior distribution rather than a point estimate. WAIC estimates the expected log point-wise predictive density ($\widehat{elppd}$) as the measurement of model performance, defined as $
\text{WAIC} = -2 \ \widehat{elppd}$,
where
\begin{eqnarray}
\widehat{elppd} =  \sum_{i=1}^{N}\left\{\log\big( \frac{1}{B}\sum_{b=1}^{B}p(\by_{i}\mid \bm{\theta}^{(b)})\big) -
V_B\big(\log p(\by_{i}\mid\bm{\theta}^{(b)})\big)\right\}.
\label{eq:WAIC}
\end{eqnarray}
Here $\by_i=(y_{i1}, \dots, y_{iJ})$; $B$ is the number of post burn-in MCMC posterior samples; $\bm{\theta}^{(b)}$ is the posterior draw of the parameter vector from the $b$-th iteration;
and $V_{B}$ represents the sample variance denoted by $V_{B}(a_b)=\frac{1}{B-1}\sum_{b=1}^{B}(a_b-\bar{a})^2$, where $\bar{a}=\sum_{b=1}^{B}a_b$. The first term within the braces in \eqref{eq:WAIC} is the log point-wise predictive density for non-missing data, which can be considered as the goodness of fit; the second term is the estimated effective number of parameters, which can be considered as the penalty term determining model complexity.  We run BAREB for a set of different $S$ values, and choose the optimal $S$ that yields the smallest WAIC.

Another challenge in implementing BAREB is to summarize a distribution over random partitions. We follow the \citet{dahl2006model} approach, and report a point estimate of the biclustering. Consider an $N\times N$ matrix $\bm{H}$, where the element $H_{i_1, i_2} = \mathbb{P}(e_{i_1}=e_{i_2}\mid data)$ represents the estimated posterior probability of patient pairs clustered together, with $\mathbb{P}(\cdot)$ being the empirical posterior mean computed based on MCMC samples. Within each MCMC iteration, the posterior sample of the patient clustering indicator $\bm{e}$ defines an $N\times N$ clustering matrix $\bm{V}^{\bm{e}}$, with the element $V^{\bm{e}}_{i_1,i_2}=I(e_{i_1}=e_{i_2})$ defined as an indicator that patient $i_1$ is clustered with patient $i_2$. With this, we propose a least-square (LS) summary for patient clustering by minimizing the Frobenius distance between $V^{\bm{e}}$ and $H$ of the posterior pairwise co-clustering probabilities, given as $\bm{e}^{LS} = \text{arg min}_{\bm{e}}\|\bm{V}^{\bm{e}}-\bm{H}\|^2$,  which is a point estimate of the patient clustering.
Conditional on $\bm{e}^{LS}$,  we extract site-level clustering $\bm{r}_s$ from MCMC iterations in which  the patient clustering indicator $\bm{e}$ is the same as $\bm{e}^{LS}$. Then we compute the LS summary of the site-level clustering $\bm{r}^{LS}_s$ for each patient level cluster $s$, through the same formulation.

The LS summary also plays a crucial role in handling the label switching problem typical to any RJMCMC implementation (\citealp{jasra2005markov}). Here, we relabel the cluster membership indicator at each iteration to match the LS summary in a post-processing step after the MCMC runs. For example, consider patient-level clustering, and let $\bm{e}^b$ be the patient cluster membership indicator drawn from the $b$th iteration. We assign the relabelling of $\bm{e}^b$ as $\bm{e}^b_{\text{new}}$, obtained by minimizing
$\text{min}_{\bm{e}^b_{\text{new}}\in\mathcal{A}(\bm{e}^b)}\text{dist}(\bm{e}^{LS} ,\bm{e}^b_{\text{new}})$. Here, the distance $\text{dist}(\cdot, \cdot)$ is defined as the number of elements that are dissimilar in the two input vectors, and $\mathcal{A}(\bm{e}^b)$ denotes the set consisting of $S^b !$ possible relabelings for $\bm{e}^b$, where $S^b$ denotes the number of patient-level clusters in the $b$th iteration. 

\section{Simulation Study} \label{sec:sim}
In this section, we conduct simulation studies to evaluate the  performance of BAREB by comparing the posterior inference to the simulation truth. Furthermore, to elucidate the advantages of using the repulsive DPP prior that encourages the linear coefficients in different biclusters to be diverse, we compared BAREB to an alternative model that uses independent priors on these linear coefficients. 

\begin{figure}[ht!]
\begin{tabular}{ccc}
\includegraphics[width=.32\textwidth]{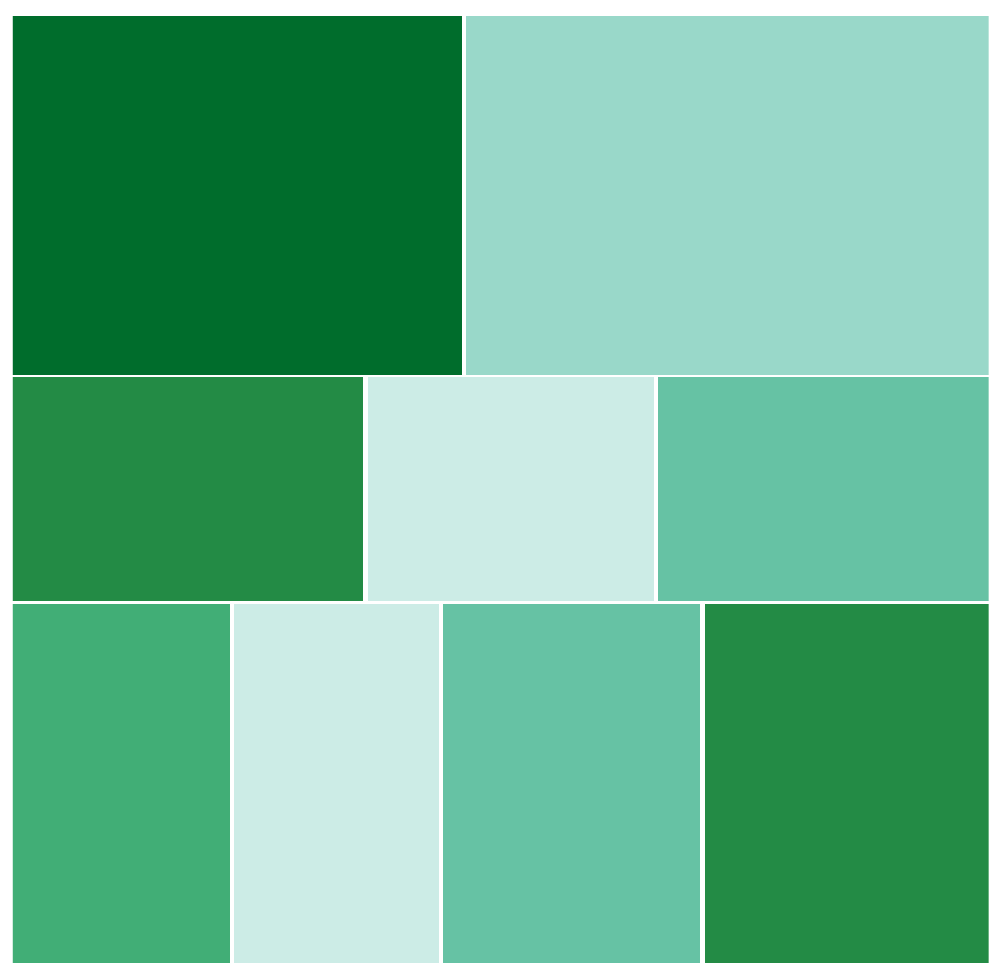}&\includegraphics[width=.32\textwidth]{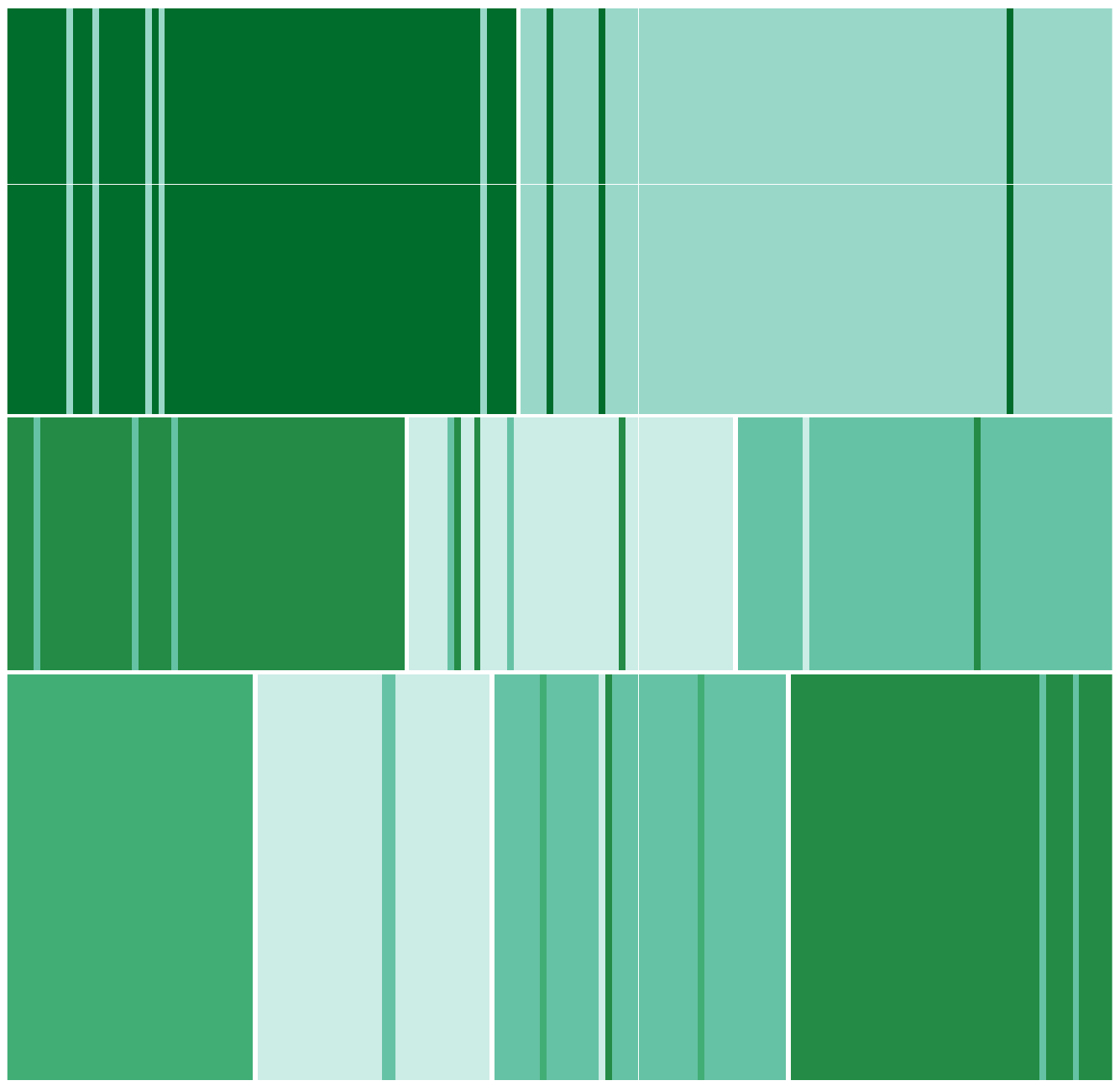}&\includegraphics[width=.32\textwidth]{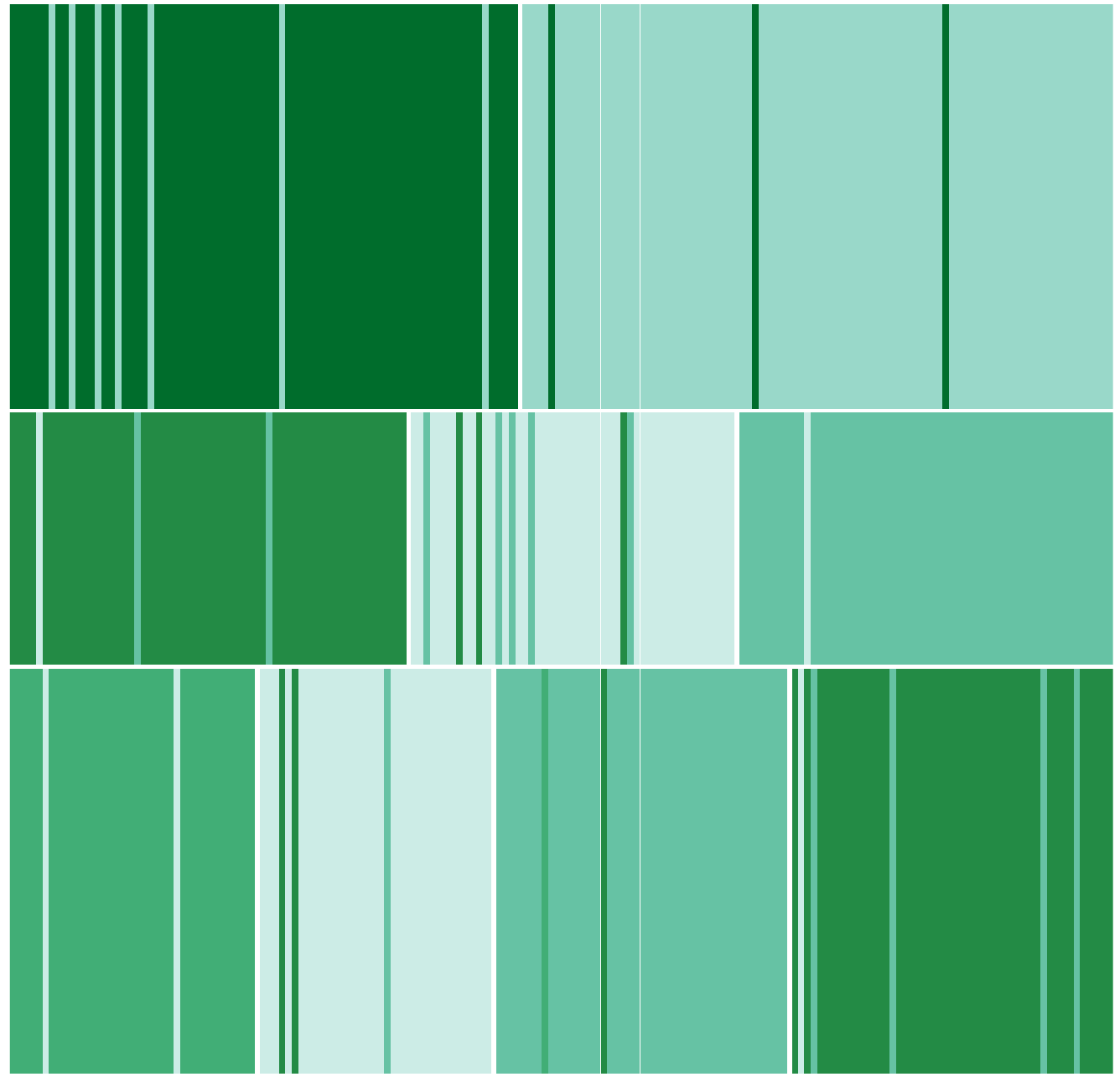}\\
(a) Simulation truth &(b) BAREB &(c) Indep\\
\end{tabular}
\caption{Simulation Illustration of biclustering: (panel a) The heatmap of biclustering in the simulation truth, (panel b) The estimated biclustering under \textit{BAREB}, (panel c) the estimated biclustering under the \textit{Indep} model. Different colors represent different biclusters. }
\label{fig:clustersim}
\end{figure}

We simulated a data matrix $Y$ with $N=80$ patients and $J=168$ tooth sites, with the true number of patient clusters $S_{0}=3$. All 80 patients have the same probability to be assigned to the three clusters. We assumed the three patient clusters partitioned the tooth sites into $(D_{10},D_{20},D_{30})=(2,3,4)$ site clusters, respectively, where each tooth site was equally assigned to the two site clusters in patient cluster 1, the three site clusters in patient cluster 2, and the four site clusters in patient cluster 3. Figure \ref{fig:clustersim}(a) illustrates the simulated true biclustering scheme, with rows representing patients, and columns representing tooth sites. We generated three patient-level covariates, $\bx_i=(x_{i1}, x_{i2}, x_{i3})$ with two continuous covariates $x_{i1}$ and $x_{i2}$ generated from $N(0, 3^2$), and one binary $x_{i3}$ generated from Bernoulli(0.5). The site-level covariate vector $\bz_j=(1, z_{j1}, z_{j2})$ was generated as $z_{j1}$ from $N(0, 3^2$), and $z_{j2}$ from Binomial(5, 0.5). For the linear coefficients $\{\widetilde{\bbeta}_{s}\}_{s=1}^{S_0}$ and $\{\widetilde{\bgamma}_{sd}\}_{d=1,s=1}^{D_{s0},\ \ S_0}$, we fixed them as in Table T1 (Supplementary Materials). Conditional on $e_i=s$ and $r_{sj}=d$, the observed response $y_{ij}$ was generated from $y_{ij}\mid e_i=s, r_{sj}=d\sim N(\bx_i\widetilde{\bbeta}_s+\bz_j\widetilde{\bgamma}_{sd} + \nu_{ij}, \sigma_0^2)$, where $\sigma_0=1$ and $\bm\nu_i$ was generated from the CAR model with parameters $\sigma_{sp}^2 = 4$ and $\rho = 0.96$. In the missing model~\eqref{eq:missing}, we assumed $\bm c = (c_0, c_1)= (0.1, 0.2)$, leading to about 20\% missing teeth in the simulated data.

\begin{figure}[ht!]
\begin{centering}
\includegraphics[scale=0.5]{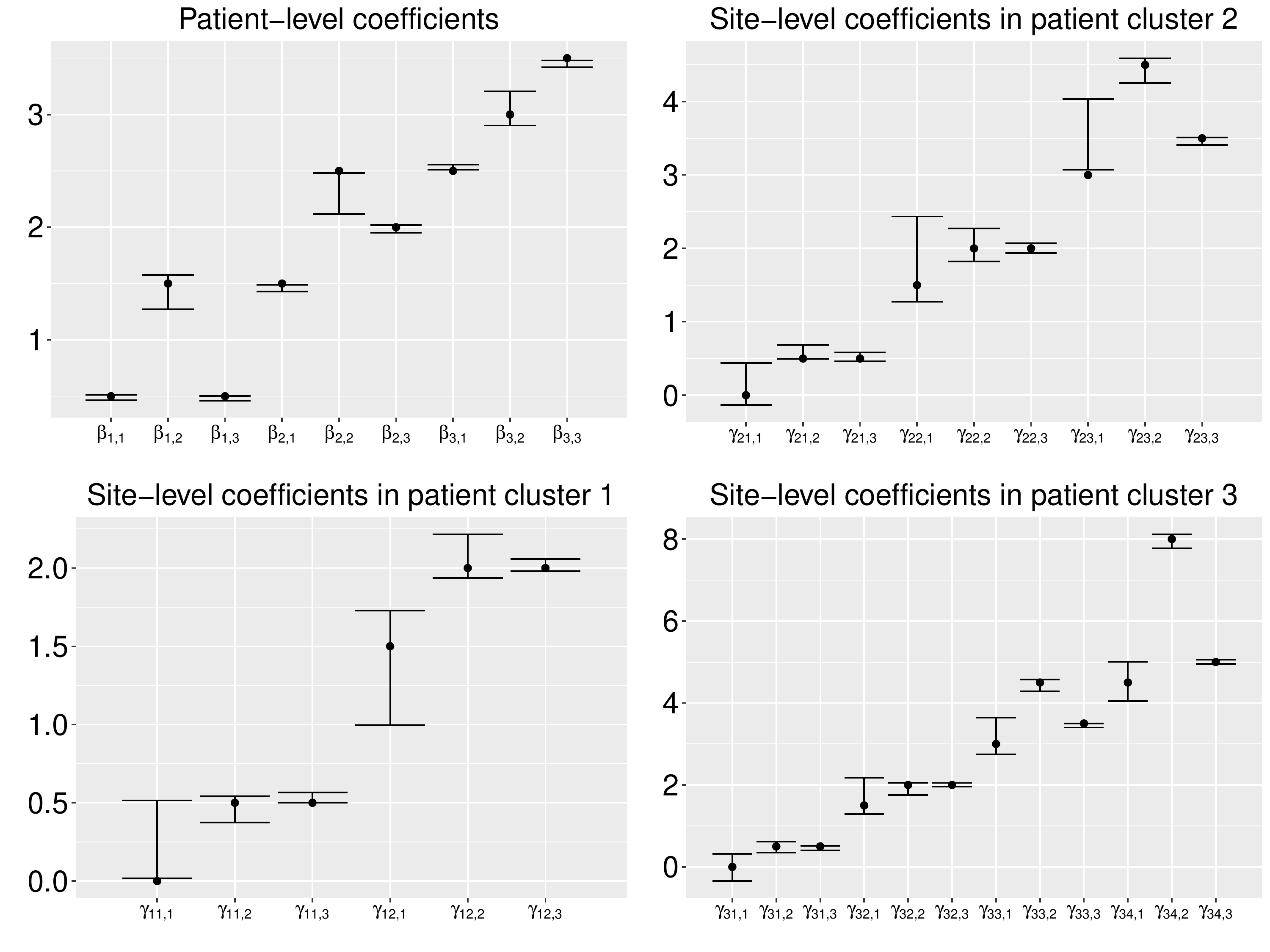}\\
\end{centering}
\caption{Simulation study: Posterior mean and 95\% CIs for the estimated $\{\widetilde{\bbeta}_s\}_{s=1}^{\hat{S}}$ and $\{\widetilde{\bgamma}_{sd}\}_{d=1,s=1}^{\hat{D}_s, \ \ \hat{S}}$, the patient- and site-level parameters under BAREB. Black dots represent the simulated true values. }
\label{fig:betagammaCI}
\end{figure}

We applied the proposed BAREB to the simulated dataset. The hyperparameters were set to be $\bm{\alpha} = (1,\dots, 1)$,  $\bm{\alpha^{\phi}_{s}} = (1,\dots, 1)$ for $s=1,\dots, S$, $a_{\sigma^2}=b_{\sigma^2}=1/2$, $\sigma^2_{\theta_{\beta}}=100$, $a_{sp}=1$, $b_{sp}=1$, $a_{\rho}=0.95$, $b_{\rho}=1$, $\sigma^2_{\theta_{\gamma}}=100$, and $\sigma_{c0}^2 = 100$.
We considered $S\in\{2,\ldots,10\}$. The RJMCMC sampler was implemented with an initial burn-in of 3000 iterations, followed by $B=2000$ post-burn-in iterations. A laptop computer with 2 GHz Intel Core i5 processor with 8 GB memory took around 45 minutes to run 5000 iterations. Convergence diagnostics assessed using \texttt{R} package \texttt{coda} revealed no issues. WAIC identified $\hat{S}=3$, which was the same as the simulation truth. The LS summary of the posterior on $\eb$ was calculated. Then, conditional on $\eb^{LS}$, we calculated the LS estimates of site clusters $\br^{LS}$. Figure \ref{fig:clustersim}(b) plots the LS summary of the posterior on $\eb$ and $\br$, showing that the BAREB correctly assigns the patients to the three patient clusters. For site clusters in each patient cluster, BAREB identified $\hat{D}_1=2, \hat{D}_2=3$, and $\hat{D}_3=4$, which also matches the simulation truth. As shown in Figure \ref{fig:clustersim}(b), BAREB assigns most of the tooth sites to their simulated true site clusters. For instance, in patient cluster 1, only four sites in site cluster 1 were misclassified; in patient cluster 3, site clusters 1 and 4 were correctly identified, while only one site in cluster 2 and two sites in cluster 3 were misclassified. Table T1 (Supplementary Materials) reports the posterior mean and the mean squared error (MSE) of the estimated coefficients, where MSE is computed as the mean squared differences between the posterior samples and the simulated true values across post-burn-in iterations. Compared to the simulation truth, BAREB can accurately estimate these coefficients with small MSE. We also plot the 95\% estimated credible intervals (CI) of the coefficients in Figure~\ref{fig:betagammaCI}, where the black dots represent the simulated true values. We observe that the 95\% CIs are centered around the simulated true values.

Next, to emphasize the advantage of the repulsive DPP prior, we conduct another study by replacing the DPP priors on $\widetilde{\bbeta}_s$'s and $\widetilde{\bgamma}_{sd}$'s with the independent multivariate normal (MVN) priors.
To be precise, for the patient-level $\widetilde{\bbeta}_s$, $s=1, \dots, S$, we assume independent MVN priors. For the site cluster membership indicator $\br_s$ in patient cluster $s$, we consider a P\'olya urn prior $p(\br_s) \propto \alpha^{D_s}_1\prod_{d=1}^{D^s}\Gamma(n_{sd})$, where $\alpha_1$ is the total mass parameter of the P\'olya urn scheme, and $n_{sd}$ is the number of tooth sites in site cluster $d$ of patient cluster $s$. Conditional on $\eb$ and $\br$, we assume independent MVN priors on $\widetilde{\bgamma}_{sd}$, $d=1, \ldots, D_s; s=1, \ldots, S$.  We coin this as the `Indep' model. Based on the WAIC criteria, Indep identified $\hat{S}=3$ patient clusters, which agrees with the truth. Figure \ref{fig:clustersim}(c) plots the LS summary of the posterior on $\eb$ and $\br$ under this Indep model. We observe that although Indep assigns patients to their simulated true patient clusters, it fails to identify tooth-site clusters within the patient clusters.

Figure F3 (Supplementary Materials) compares the posterior histograms of the number of site-clusters within each patient cluster under the BAREB and Indep models. Clearly, BAREB recovers the ground truth, while the Indep overestimates the number of site clusters with substantial probability, which is a well-known phenomenon of applying Bayesian nonparametric priors to clustering (\citealp{xu2013nonparametric}). Thus, we observe that the DPP is advantageous over independent priors in biclustering scenarios.

\section{Application: GAAD Data}\label{sec:apply}
The GAAD study (\citealp{fernandes2009periodontal}) was primarily designed to explore the relationship between PD and diabetes status, determined by the glycosylated haemoglobin (HbA1c) level. Excluding patients with all teeth missing, we have $N=288$ patients in the dataset. We considered several patient-level covariates as potential risk factors of PD, which includes Age (in years), Gender (female=1, male=0), Smoking indicator (smoker=1, non-smoker=0), and HbA1c (high level=1, controlled=0). We also considered a site-level covariate, the jaw indicator (upper jaw=1, low jaw=0). Table T2 (Supplementary Materials) lists the patient characteristics from the dataset.

We applied BAREB to the PD dataset, considering $S=\{2,\cdots,10\}$. The hyperparameters were set to be $\bm{\alpha} = (1,\ldots, 1)$,  $\bm{\alpha^{\phi}_{s}} = (1,\ldots, 1)$ for $s=1,\ldots, S$, $a_{\sigma^2}=b_{\sigma^2}=1/2$, $\sigma^2_{\theta_{\beta}}=100$, $a_{sp}=1$, $b_{sp}=1$, $a_{\rho}=0.8$, $b_{\rho}=1$, $\sigma^2_{\theta_{\gamma}}=100$, and $\sigma_{c0}^2 = 100$. For each $S$, we used 5,000 post burn-in samples after 10,000 iterations to compute our posterior estimates. WAIC identified $\hat{S}=4$ patient clusters. We then computed the LS estimates $\eb^{LS}$ and $\br^{LS}$ to summarize the posterior inference for biclustering. The four patient clusters had cluster sizes of 3, 174, 80, and 31 patients, respectively. The numbers of site clusters within patient clusters are 3, 2, 2, and 1.

\begin{figure}[ht!]
\begin{center}
\begin{tabularx}{\linewidth}{c X}
&
 \hfill \makecell{ \includegraphics[scale=0.18]{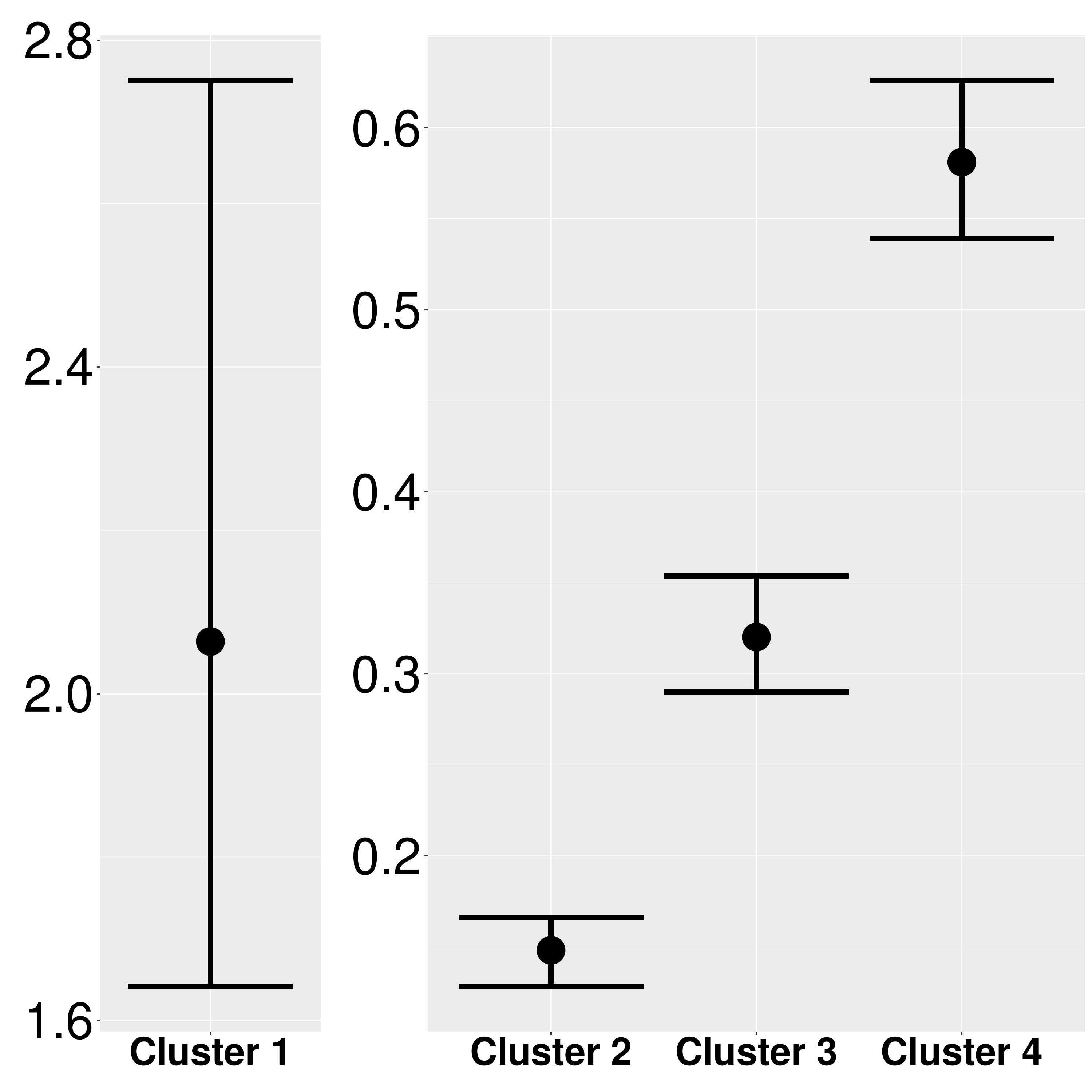}\\ (a) Age}
 \hfill \makecell{ \includegraphics[scale=0.18]{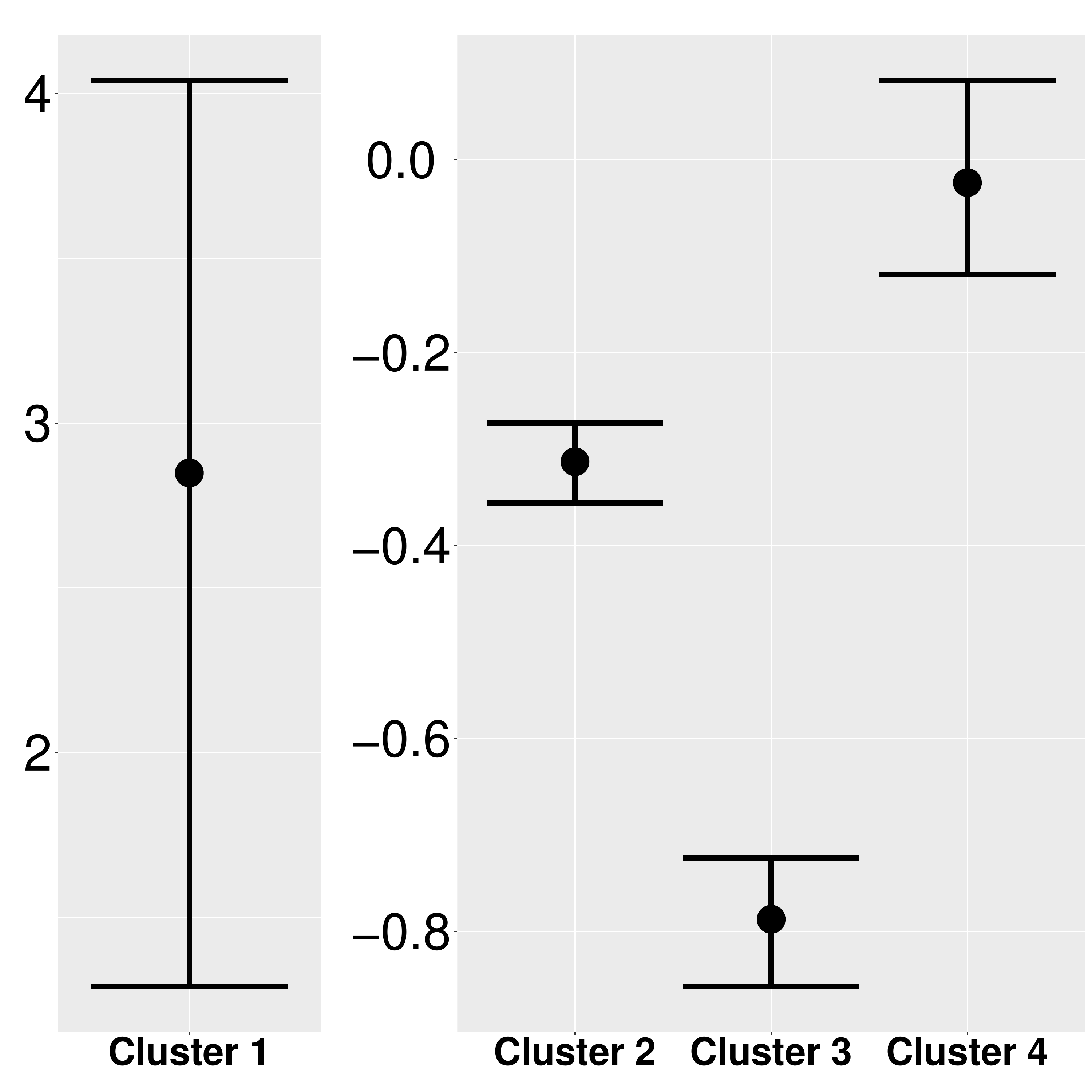}\\ (b) Gender} \hfill\null \\
&
\hfill \makecell{ \includegraphics[scale=0.18]{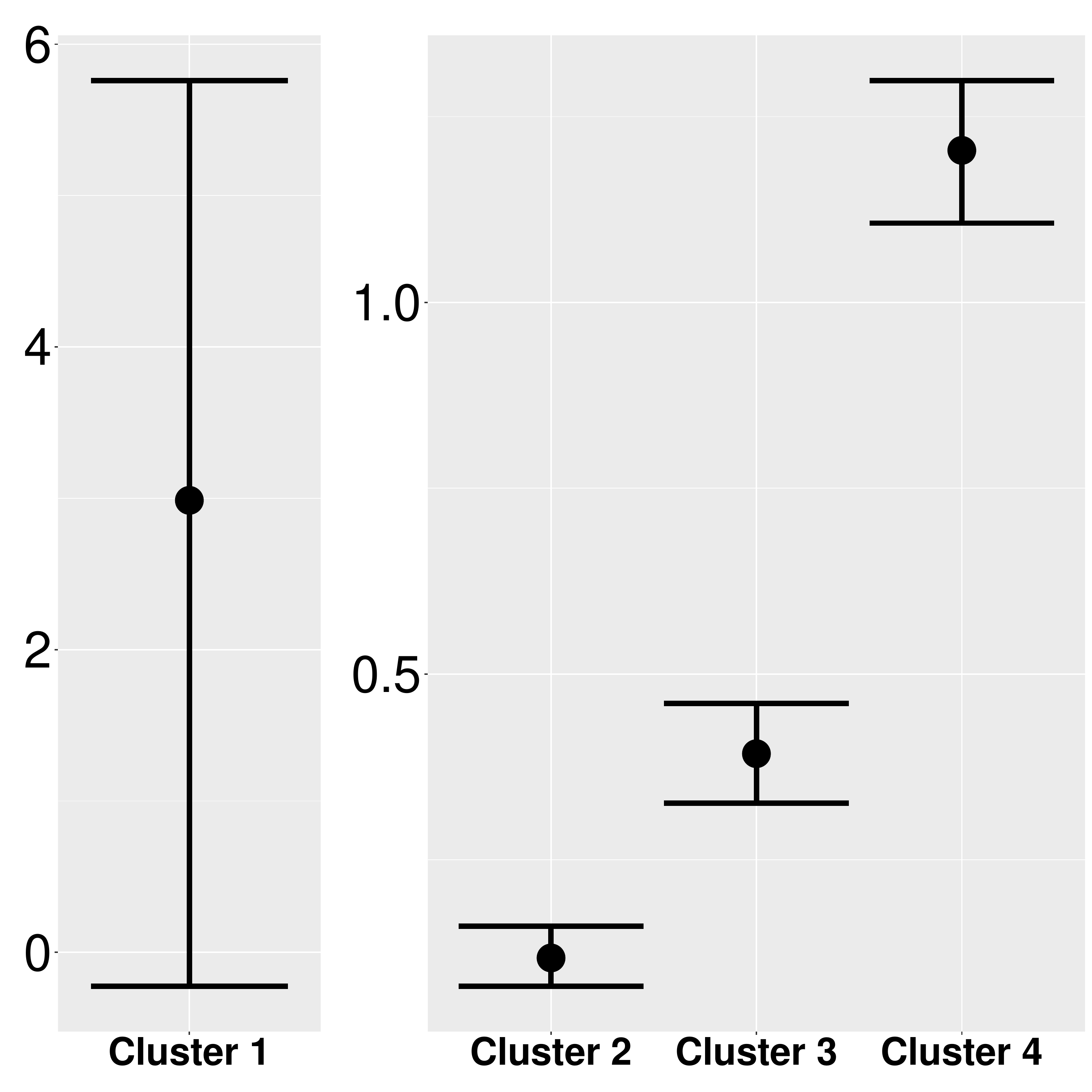}\\ (c) Smoking indicator}
\hfill \makecell{ \includegraphics[scale=0.18]{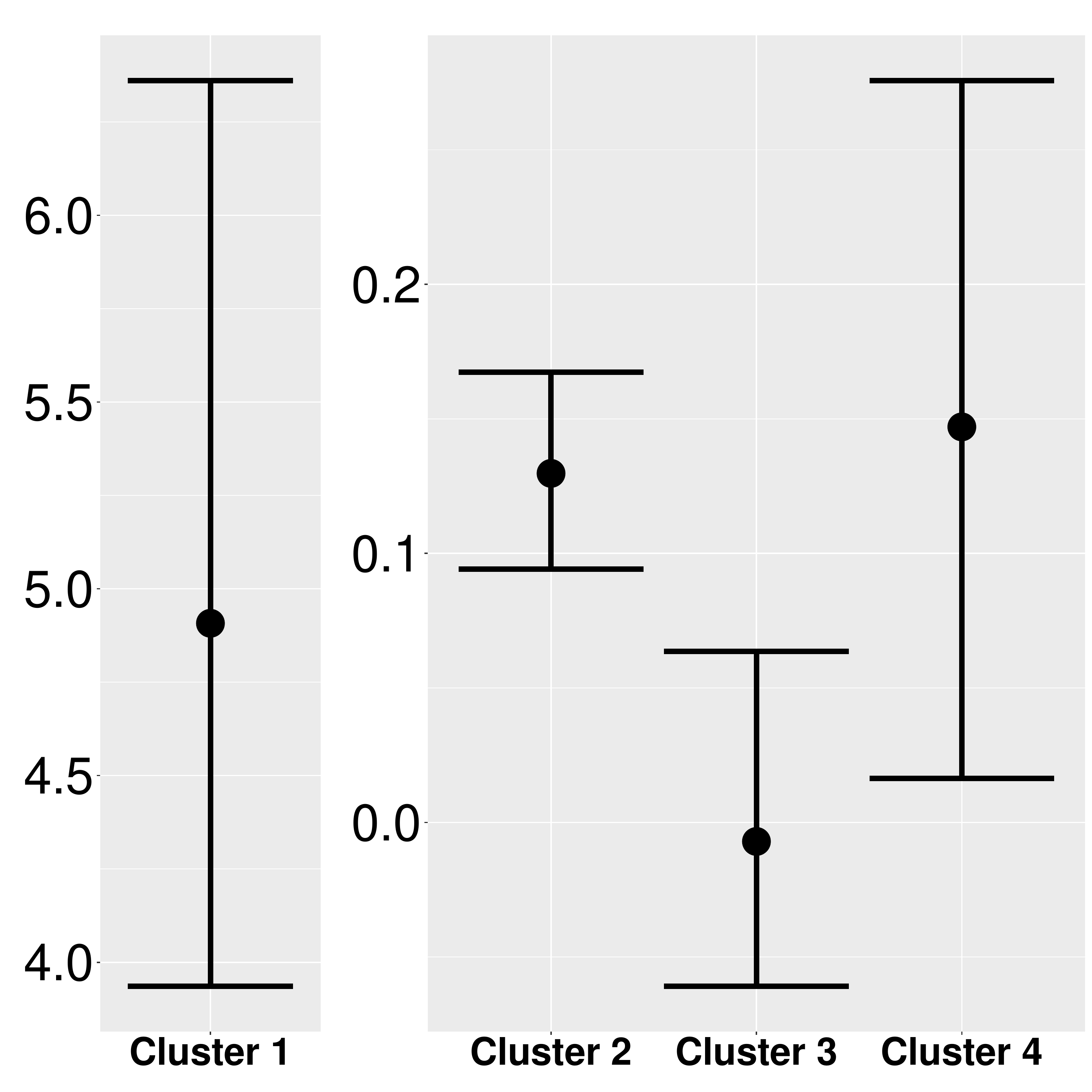}\\ (d) HbA1c} \hfill\null \\
&
\hfill \makecell{ \includegraphics[scale=0.18]{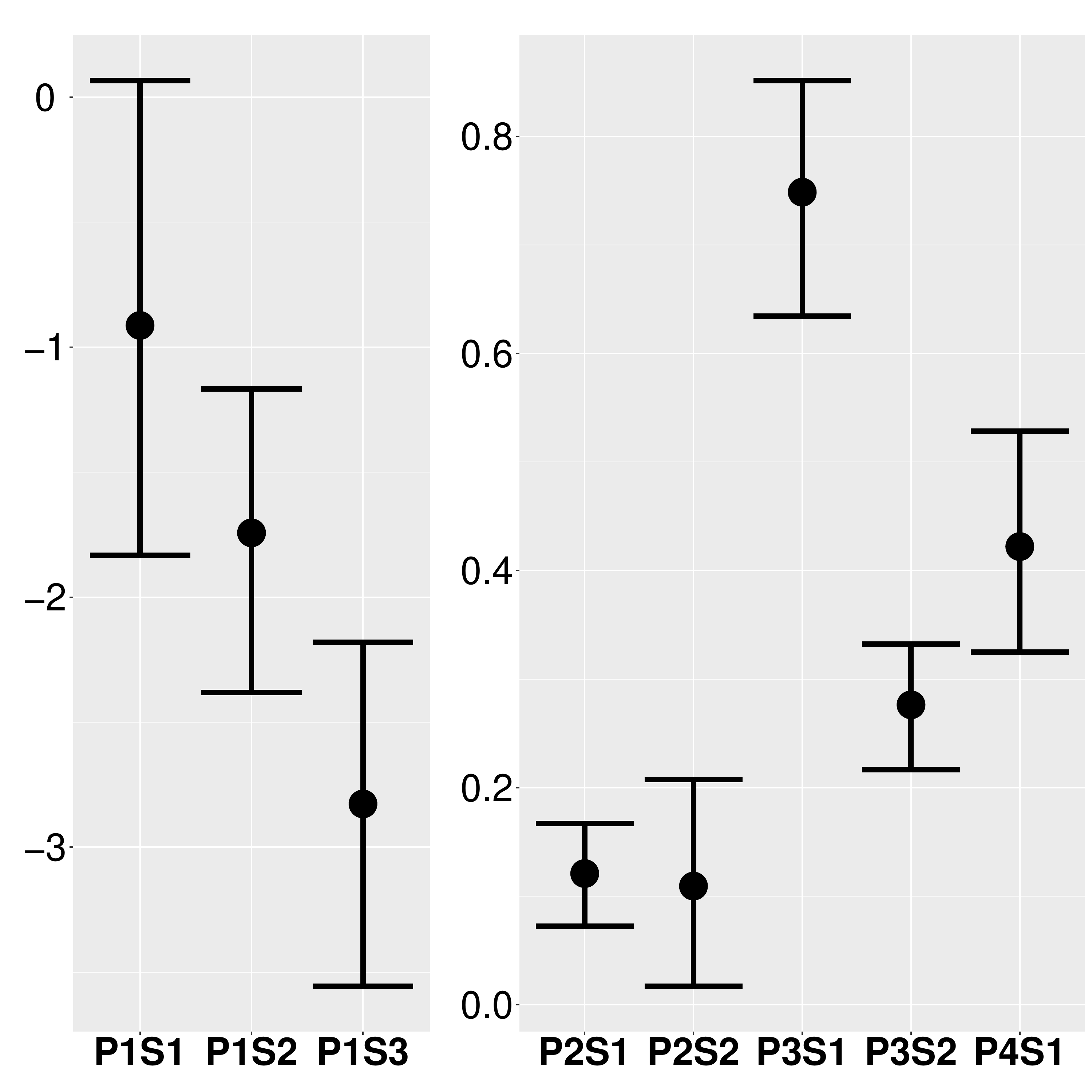}\\ (e) Jaw Indicator} \hfill\null \\
\end{tabularx}
\end{center}
\caption{Posterior mean and 95\% credible intervals for the parameters corresponding to the patient-level covariates: (a) Age, (b)Gender, (c) Smoking status indicator, (d) HbA1c, and site-level covariate (e) Jaw indicator. In panel (e), P1S1 denotes site cluster 1 in patient cluster 1,  P1S2 denotes site cluster 2 in patient cluster 1, etc. }
\label{fig:CIBetaGamma}
\end{figure}

Figure~\ref{fig:CIBetaGamma} summarizes the posterior mean and 95\% credible intervals of the covariates within the four estimated patient clusters. Note, we separately plot the parameters for patient cluster 1 for visualization, since its scale is very different from the other clusters. Also, due to the small cluster size ($n_1=3$) in patient cluster 1, the estimated parameters may not be very reliable, with the 95\% CIs much wider than those observed in other patient clusters, as expected. Further looking into patient cluster 1 reveals these 4 patients have at least half of the teeth missing, with an average missingness rate being 69\%. Therefore, we excluded the patient cluster 1 from the following clinical interpretations. As shown in Figure~\ref{fig:CIBetaGamma}, the effects of the patient-level covariates are quite distinct among different patient clusters.
Panel (a) shows periodontal health deteriorates with age, since CIs of Age in all clusters are positive, and exclude 0. Similarly, a significant positive association was found between PD and smoking, a factor that has been believed to increase the risk of PD (\citealp{leite2018effect}). Panel (b) shows that males are more likely to have severe PD than females, confirming previous findings (\citealp{mamai2016gender}), while panel (c) reveals smoking to be an important determinant of PD, mostly for patient clusters 3 and 4.
From panel (d), we observe that PD has a positive correlation with diabetes (HbA1c),
revealing that diabetes is possibly an important risk factor of PD (\citealp{jansson2006type}).
The plots for the site-level Jaw Indicator in panel (e) implies that teeth in the upper jaw (maxilla) are more likely to develop PD. Although this finding is inconclusive, previous studies (\citealp{shigli2009relative, volchansky2016patterns}) seem to indicate that some maxillary teeth, such as the left and right central incisors and the first premolar, experience a higher rate of missingness (due to PD), than the mandibular (lower jaw) teeth. In the GAAD dataset, the tooth missingness rates in the maxilla and mandible are, respectively,  37\% and 28\%. Among the non-missing teeth, the mean CAL values (combining all teeth) are 2.03 and 1.83, respectively, for the maxilla and mandible.

\begin{figure}[ht!]
\begin{center}
\begin{tabularx}{\linewidth}{c X}
&
    \hfill \makecell{ \includegraphics[width=6cm, height=5cm]{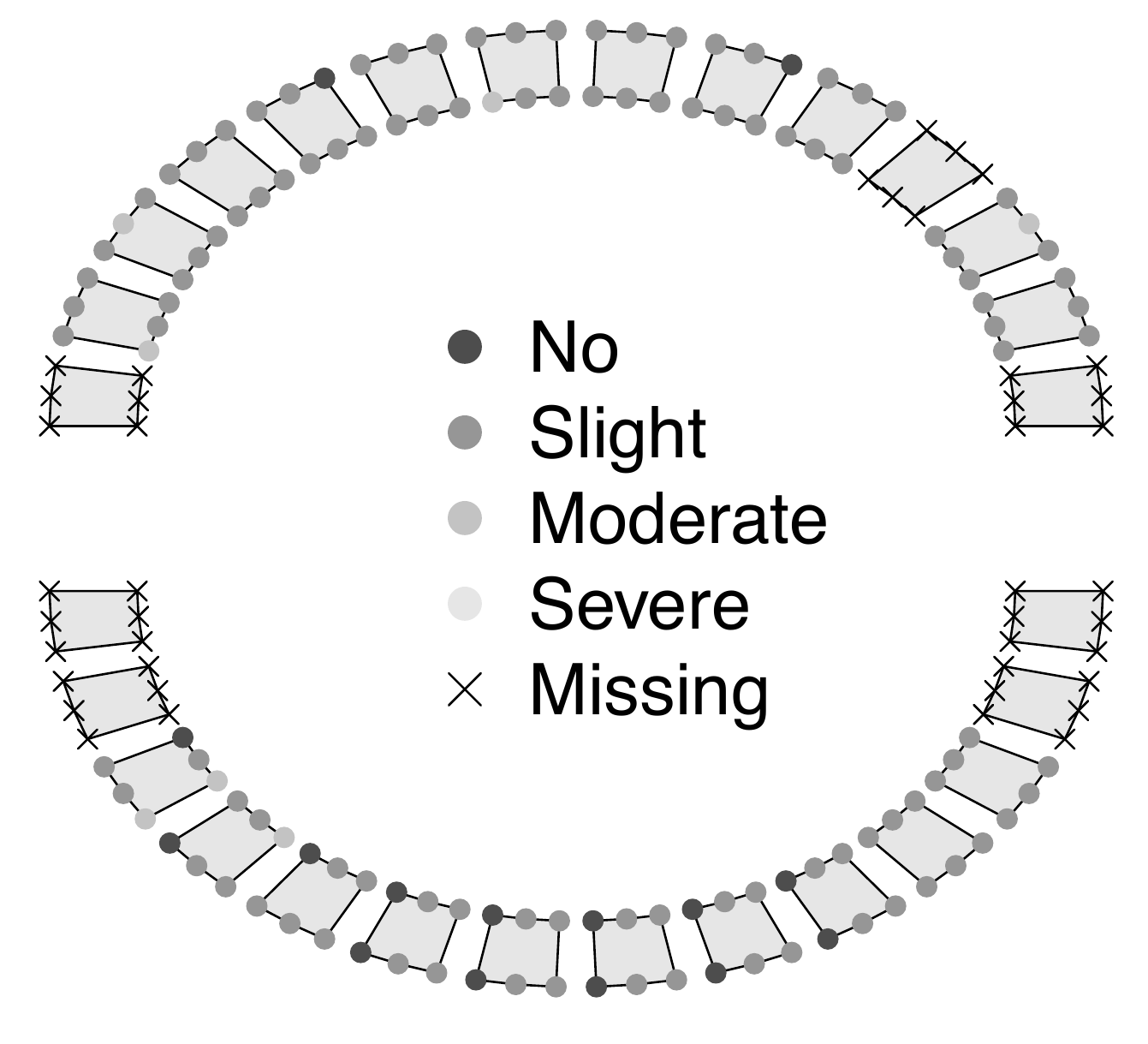}\\ (a) Patient cluster 2}
     \hfill \makecell{ \includegraphics[width=6cm, height=5cm]{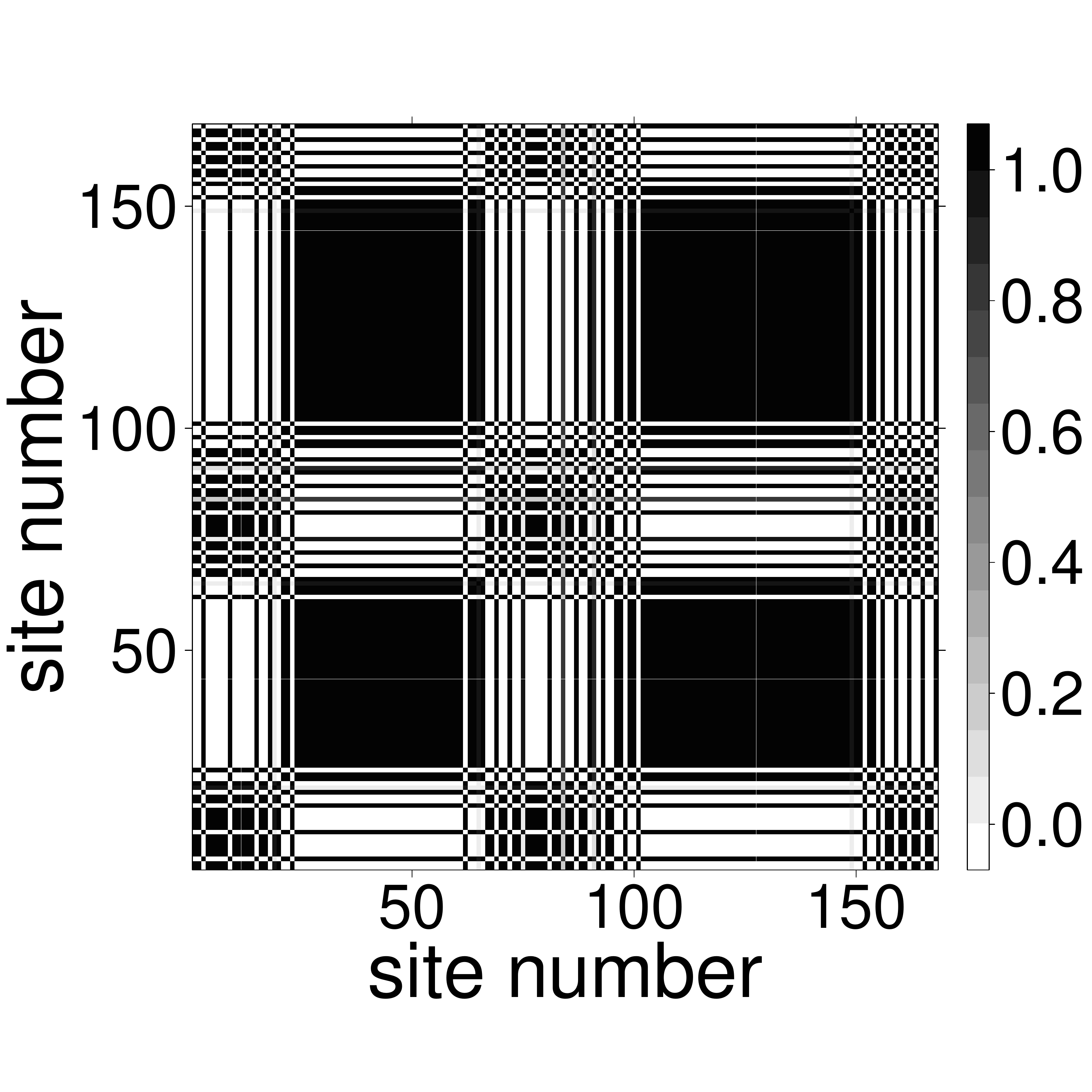}\\ (b) Patient cluster 2} \hfill\null \\
     &
    \hfill \makecell{ \includegraphics[width=6cm, height=5cm]{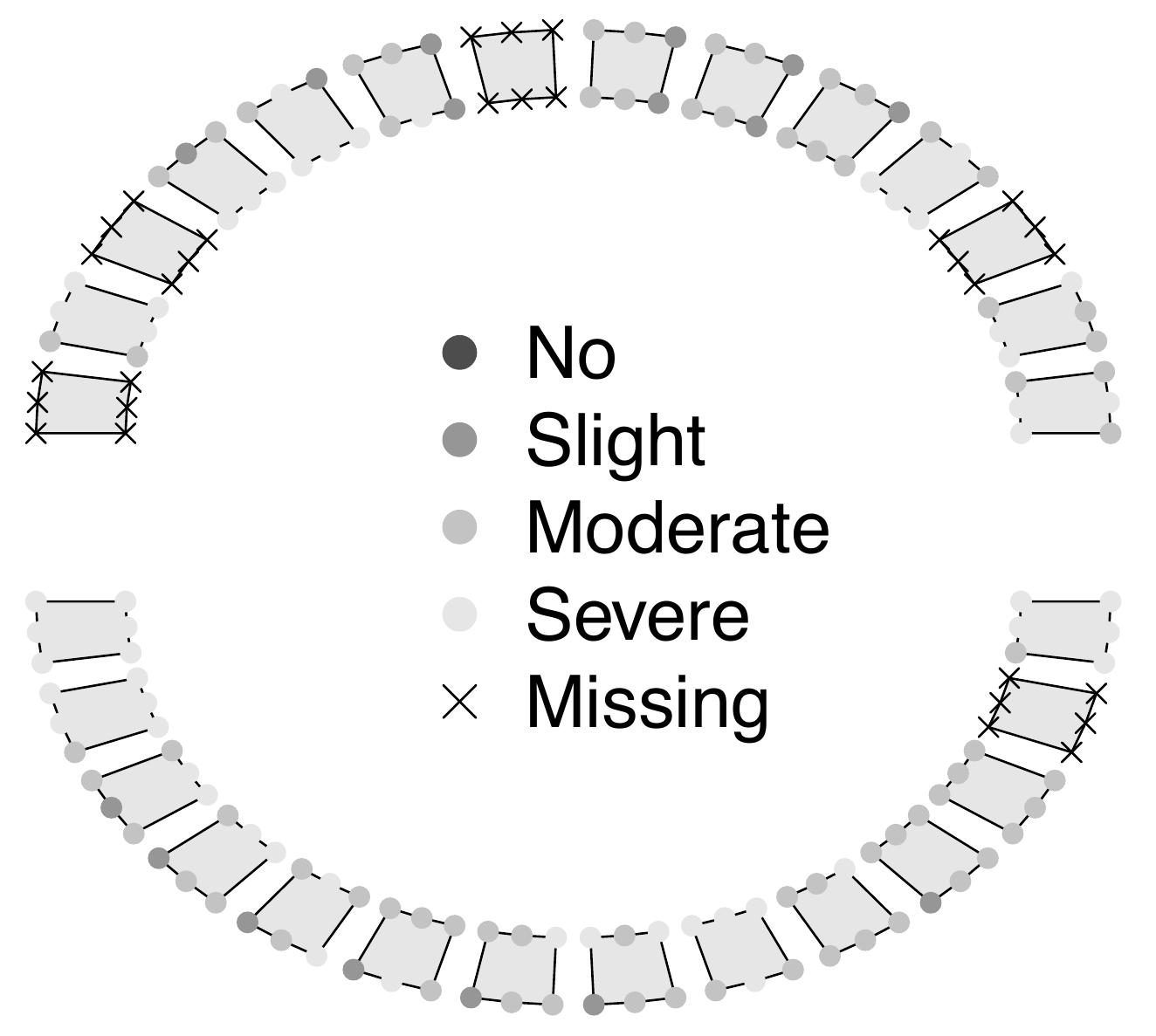}\\ (c) Patient cluster 3}
     \hfill \makecell{ \includegraphics[width=6cm, height=5cm]{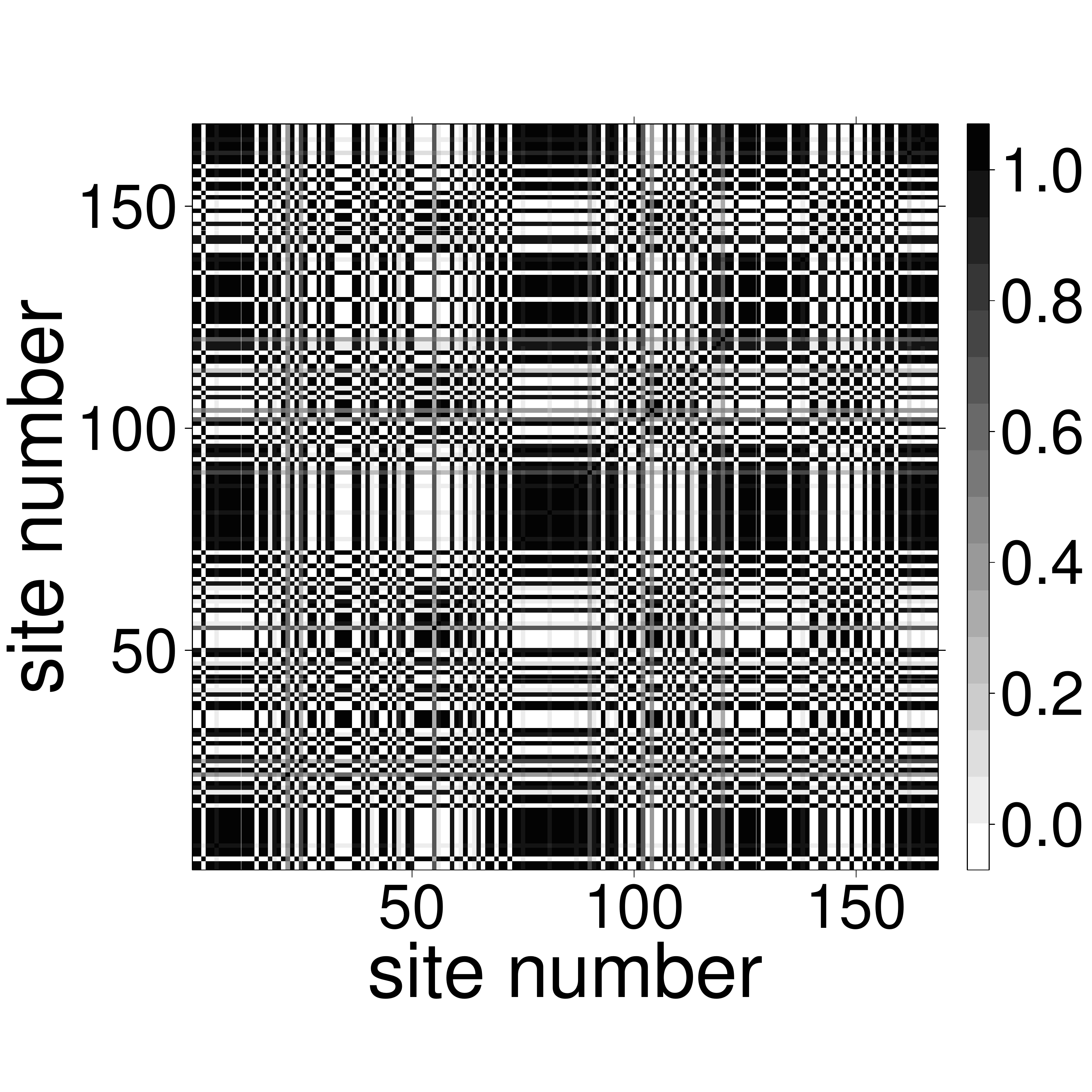}\\ (d) Patient cluster 3} \hfill\null \\
\end{tabularx}
\end{center}
\caption{Site-level clustering in GAAD data. Panels (a) and (b) plot the ordinal CAL levels, and the heatmap of the posterior clustering probability of tooth-sites for a randomly selected patient from patient cluster 2. Panels (c) and (d) plot the same, respectively, now for a randomly selected patient from patient cluster 3.}
\label{fig:clusters}
\end{figure}

Next, we report site clustering in patient clusters 2 and 3. Patient cluster 1 is excluded owing to its small cluster size, while patient cluster 4 is excluded as there is only one site cluster from the LS estimate. Figure~\ref{fig:clusters} (panels a and c) display the ordinal CAL values from two randomly selected patients in patient clusters 2 and 3, respectively, based on the classification of the American Association of Periodontology (\citealp{armitage1999development}). The five ordinal categories are (i) no PD (CAL 0-1 mm), (ii) slight PD (CAL 1-2 mm), (iii) moderate PD (CAL 3-4 mm), (iv) severe PD (CAL 5 $\ge$ 5mm), and (v) missing (if the tooth is missing).  Figure~\ref{fig:clusters} (panels b and d) plot the heatmaps of the estimated posterior probability of each pair of sites being clustered together, as described in subsection~\ref{s:post}. BAREB estimates distinct site clustering patterns in different patient clusters. In particular, for the randomly selected patient in patient cluster 2, two maxillary molars and four mandibular molars are missing. From the corner dark squares in Figure~\ref{fig:clusters} (panel b), these missing molar sites are clustered together. For example, the missing maxillary molar sites 1-6 and 79-84 are clustered, along with the mandibular missing sites 157-168. Also, the four rectangular black patches reveal tooth sites of the same type are more likely to be clustered together. On the contrary, for the randomly selected patient from patient cluster 3, the corresponding plot (panel d) produces a checkerboard pattern, where the missing sites at different site locations tend to cluster with a high probability.

\begin{figure}[ht!]
\begin{centering}
\includegraphics[scale=0.4]{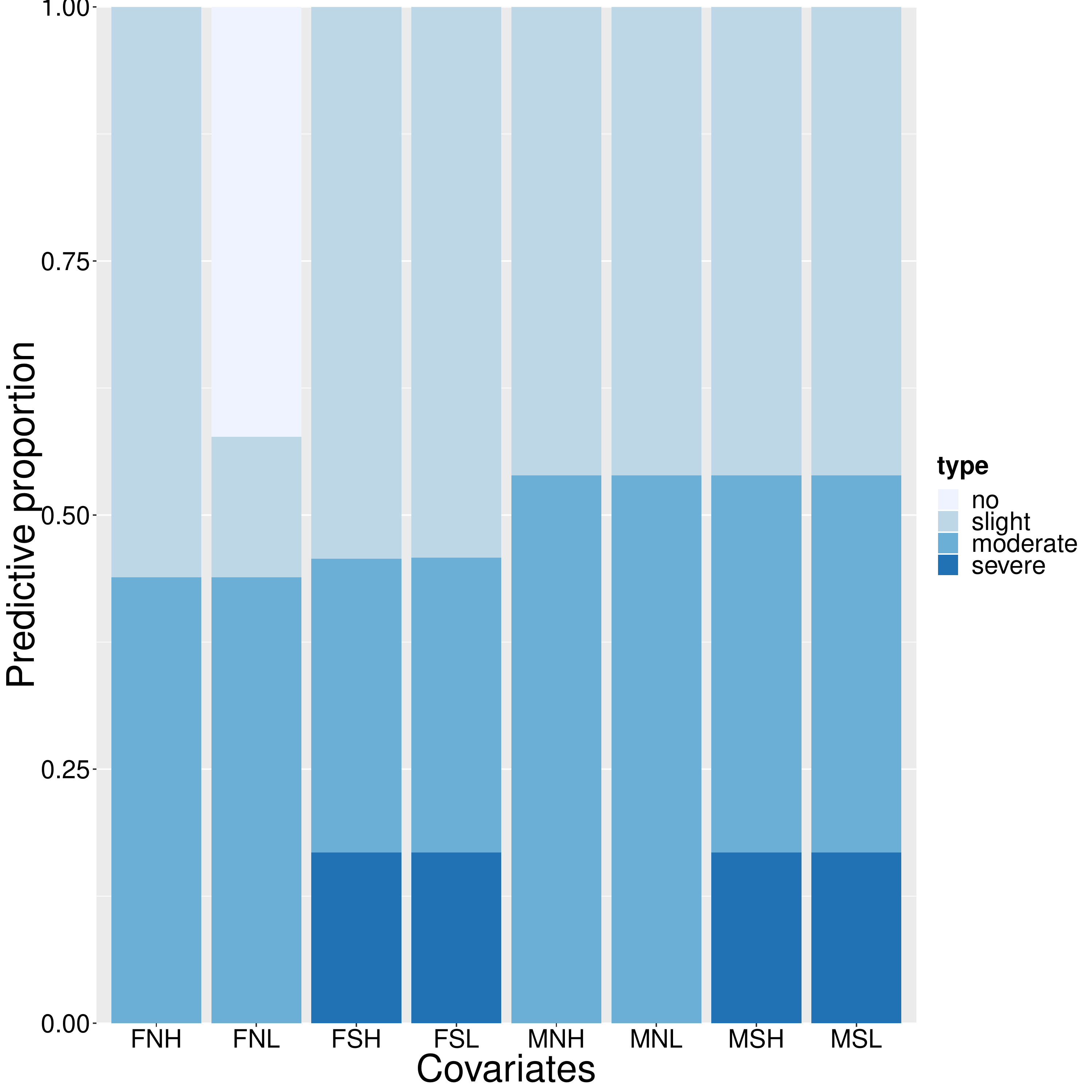}\\
\end{centering}
\caption{Posterior predictive probabilities of ordinal CAL categories for site \# 120 of a mandibular incisor from a hypothetical patient with age 55.27 years, under various combinations of gender, smoking and HbA1c levels.}
\label{fig:predictive}
\end{figure}

Finally, in Figure~\ref{fig:predictive}, we report the posterior predictive probabilities of the aforementioned ordinal CAL categories.  For illustration, we consider a specific tooth-site (\# 120) from the mandibular incisor of a hypothetical patient with mean age 55.27 years old, under all possible combinations of gender, smoking and HbA1c levels, where F, M, N, S, L, and H denote female, male, non-smoker, smoker, controlled HbA1c, and high HbA1c, respectively. Considering a tooth-site with no, or slight PD as `healthy', the probability being healthy is 0.561 for a female non-smoker with controlled HbA1c, while the probability is 0 for a male smoker with high HbA1c. Specifically, females have higher probability of having healthy teeth than males (for example, FNL = 0.561, versus MNL = 0.461); smokers have lower probability than nonsmokers (FSL = 0.542 versus FNL = 0.561); controlled HbA1c have higher probability of having no PD than high HbA1c  (FNL = 0.423 versus MNH = 0.000), and so on. Figure F4 (Supplementary Materials) presents the density histogram plot of the CAL response, overlaid with the fitted curve generated from the marginal posteriors and Silverman's rule-of-thumb smoothing bandwidth. The proposed BAREB model induces a marginal density capable of accommodating the possible non-Gaussian features of the CAL. Thus, in lieu of non-Gaussian parametric (say, skew-$t$), and semiparametric assumptions for the error term $\epsilon_{ij}$ that may induce computational issues, we consider our $N(0, \sigma^2)$ assumption to be adequate. We also present the posterior means and 95\% credible intervals for other remaining parameters in Table T3 (Supplementary Materials). It is worth noting that estimate of $c_1$ is positive and significant, which unsurprisingly implies that higher values of CAL leads to higher probability of missing teeth. The estimate of $\rho$, the spatial association parameter in the CAR model, is 0.843, which corresponds to moderate spatial correlation based on the calibration of $\rho$ (\citealp{carlin2014hierarchical}). 

\section{Conclusion} \label{sec:end}
Our proposed BAREB can detect simultaneous clustering patterns among PD study patients and tooth sites, factoring in both patient- and site-level covariates, the spatial dependence among tooth sites, and possible nonrandom missingness patterns. That way, our proposal improves upon available clustering techniques into learning the heterogeneity in PD among patients with distinct disease patterns. We also demonstrate the advantages of using the DPP prior over independent priors for quantifying diversity among various biclusters, balancing parsimony and interpretation. In addition, BAREB is readily implementable via \texttt{R}, and can be a welcome addition to a user's toolbox.

Although motivated from an oral health application, BAREB provides a general framework for inference on biclustering in many other applications involving a data matrix and covariates. For example, in gene expression data (where rows and columns represent genes and patient samples, respectively), BAREB can discover functionally related genes under different subsets of patients, factoring in various clinical covariates.  Furthermore, the choice of a parametric selection model to tackle non-random missingness was also due to computational reasons, leaving us no opportunity to introduce sensitivity parameters for assessing non-random missingness (\citealp{daniels2008missing}). All these are viable areas for future research, and will be pursued elsewhere.

\section{Supplementary Materials}\label{sec:ack}
Web appendices and \texttt{R/C++} codes for implementing BAREB are available with this article at the Biostatistics website, and at the \texttt{GitHub} link:
\url{https://github.com/YanxunXu/BAREB}, respectively.

\section*{Acknowledgements}
The authors thank the Center for Oral Health Research at the Medical University of South Carolina for providing the GAAD dataset. The work of Bandyopadhyay was supported by NIH grant R01-DE024984. The work of Xu was supported by NSF grant 1918854.

\bibliographystyle{apalike}
\bibliography{biclustering}

\newpage

\section*{\textbf{Web Appendix A: Figures \& Tables}}

\renewcommand{\theequation}{A-\arabic{equation}}
\setcounter{equation}{0}  

\begin{figure}[ht!]
\begin{centering}
\includegraphics[scale=0.6]{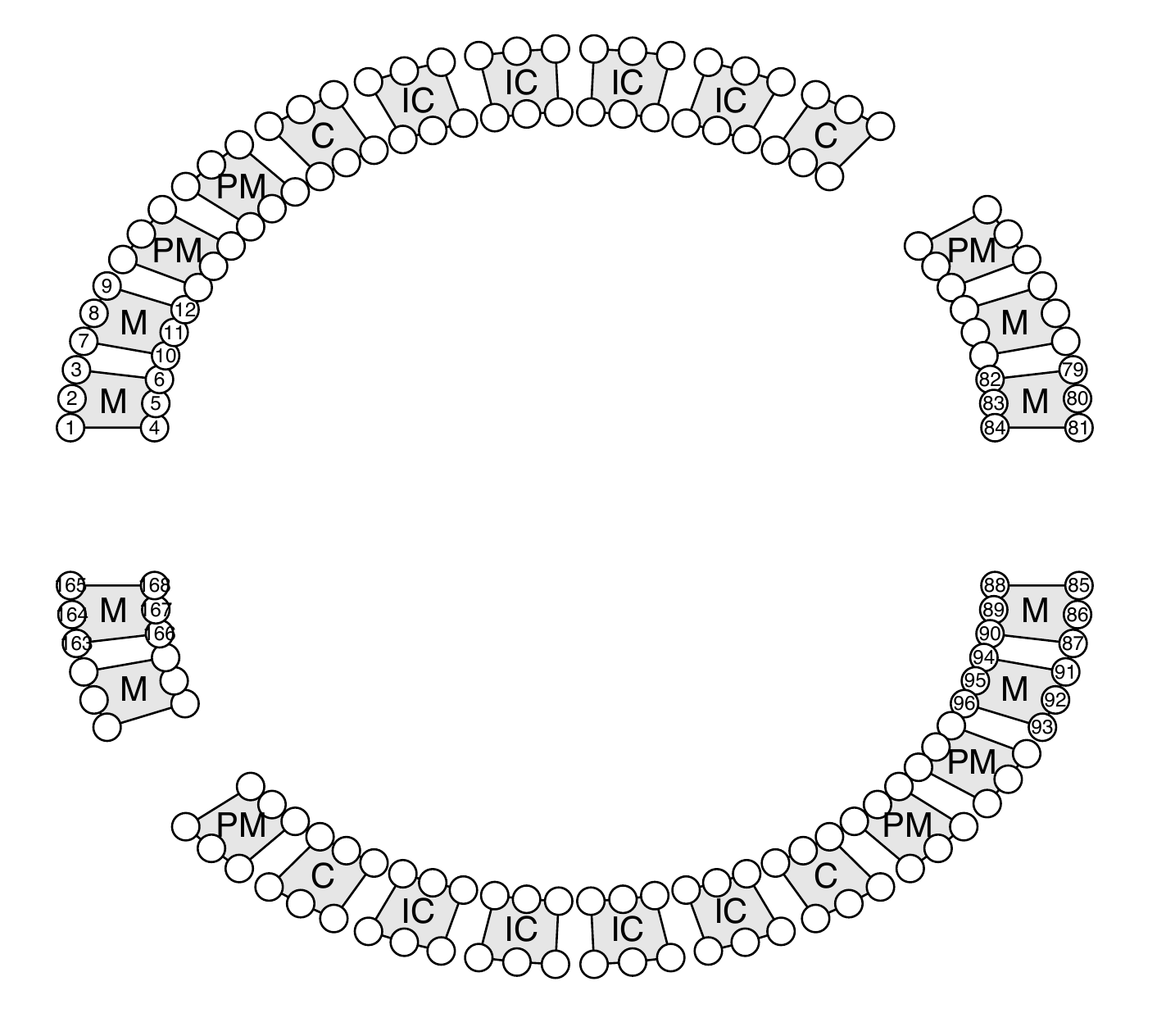}\\
\end{centering}
\captionsetup{labelformat=empty}
\caption{Figure F1: Tooth types (M = molar; PM = premolar; C = canine and IC = incisor) and site numbering (such as 1-6 for the second molar on the upper left quadrant, 7-12, etc), for a hypothetical subject with missing premolars in the upper and lower jaws.}
\label{fig:tooth}
\end{figure}

%


\begin{figure}[ht]
\begin{center}
\begin{picture}(275,130)
\put(10,36){{\Large $\bullet$}} 
\put(60,36){{\textcolor{red}{\Large $\bullet$}}} 
\put(110,36){{\Large $\bullet$}} 
\put(160,36){{\Large $\bullet$}} 
\put(210,36){{\textcolor{red}{\Large $\bullet$}}} 
\put(260,36){{\Large $\bullet$}} 
\put(10,116){{\Large $\bullet$}} 
\put(60,116){{\textcolor{red}{\Large $\bullet$}}} 
\put(110,116){{\Large $\bullet$}} 
\put(160,116){{\Large $\bullet$}} 
\put(210,116){{\textcolor{red}{\Large $\bullet$}}} 
\put(260,116){{\Large $\bullet$}} 
\put(12,50){\line(0,1){5}}\put(12,60){\line(0,1){5}}
\put(12,70){\line(0,1){5}}\put(12,80){\line(0,1){5}}
\put(12,90){\line(0,1){5}}\put(12,100){\line(0,1){5}}
\put(12,110){\line(0,1){5}}\put(12,40){\line(0,1){5}}
\put(112,50){\line(0,1){5}}\put(112,60){\line(0,1){5}}
\put(112,70){\line(0,1){5}}\put(112,80){\line(0,1){5}}
\put(112,90){\line(0,1){5}}\put(112,100){\line(0,1){5}}
\put(112,110){\line(0,1){5}}\put(112,40){\line(0,1){5}}
\put(162,50){\line(0,1){5}}\put(162,60){\line(0,1){5}}
\put(162,70){\line(0,1){5}}\put(162,80){\line(0,1){5}}
\put(162,90){\line(0,1){5}}\put(162,100){\line(0,1){5}}
\put(162,110){\line(0,1){5}}\put(162,40){\line(0,1){5}}
\put(262,50){\line(0,1){5}}\put(262,60){\line(0,1){5}}
\put(262,70){\line(0,1){5}}\put(262,80){\line(0,1){5}}
\put(262,90){\line(0,1){5}}\put(262,100){\line(0,1){5}}
\put(262,110){\line(0,1){5}}\put(262,40){\line(0,1){5}}

\linethickness{2pt}
\put(10,40){\line(1,0){100}} \put(160,40){\line(1,0){100}}
\put(10,120){\line(1,0){100}} \put(160,120){\line(1,0){100}}
\put(110,40){\line(1,0){5}}\put(120,40){\line(1,0){5}}
\put(130,40){\line(1,0){5}}\put(140,40){\line(1,0){5}}
\put(150,40){\line(1,0){5}}\put(160,40){\line(1,0){5}}
\put(110,120){\line(1,0){5}}\put(120,120){\line(1,0){5}}
\put(130,120){\line(1,0){5}}\put(140,120){\line(1,0){5}}
\put(150,120){\line(1,0){5}}\put(160,120){\line(1,0){5}}
\put(25,50){\line(1,0){70}} \put(25,50){\line(0,1){60}}
\put(95,110){\line(-1,0){70}} \put(95,110){\line(0,-1){60}}
\put(175,50){\line(1,0){70}} \put(175,50){\line(0,1){60}}
\put(245,110){\line(-1,0){70}} \put(245,110){\line(0,-1){60}}
\put(60,20){\line(1,0){45}}\put(120,20){Type I}
\put(60,0){\line(1,0){5}}\put(70,0){\line(1,0){5}}
\put(80,0){\line(1,0){5}}\put(90,0){\line(1,0){5}}
\put(100,0){\line(1,0){5}}\put(120,0){Type II}
\linethickness{.5pt}
\put(200,20){Type III}\put(260,20){\line(1,0){5}}
\put(270,20){\line(1,0){5}}\put(280,20){\line(1,0){5}}
\put(290,20){\line(1,0){5}} \put(250,20){\line(1,0){5}}
\end{picture}
\captionsetup{labelformat=empty}
\caption{Figure F2: Periodontal grid for two adjacent teeth located in the same jaw. The red dots denote non-gap sites, while the black dots denote sites located in a gap. Type I-III denotes neighbors of various kinds. \label{periogrid}}
\end{center}
\end{figure}
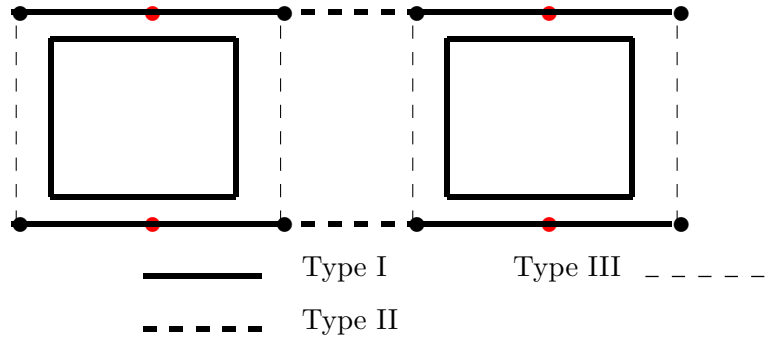

\newpage 

\begin{figure}[ht!]
\begin{tabular}{ccc}
\includegraphics[width=.33\textwidth]{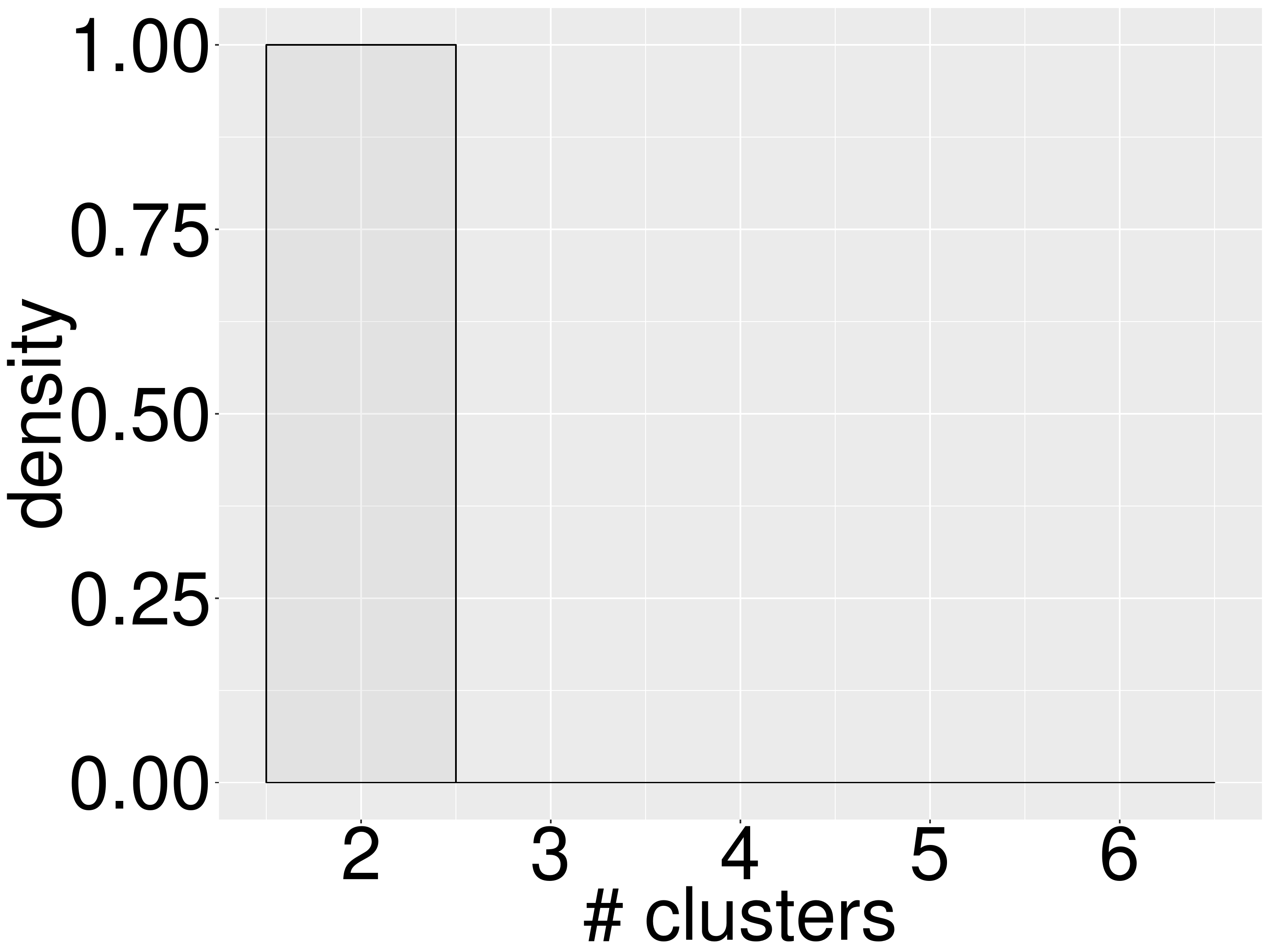}&\includegraphics[width=.33\textwidth]{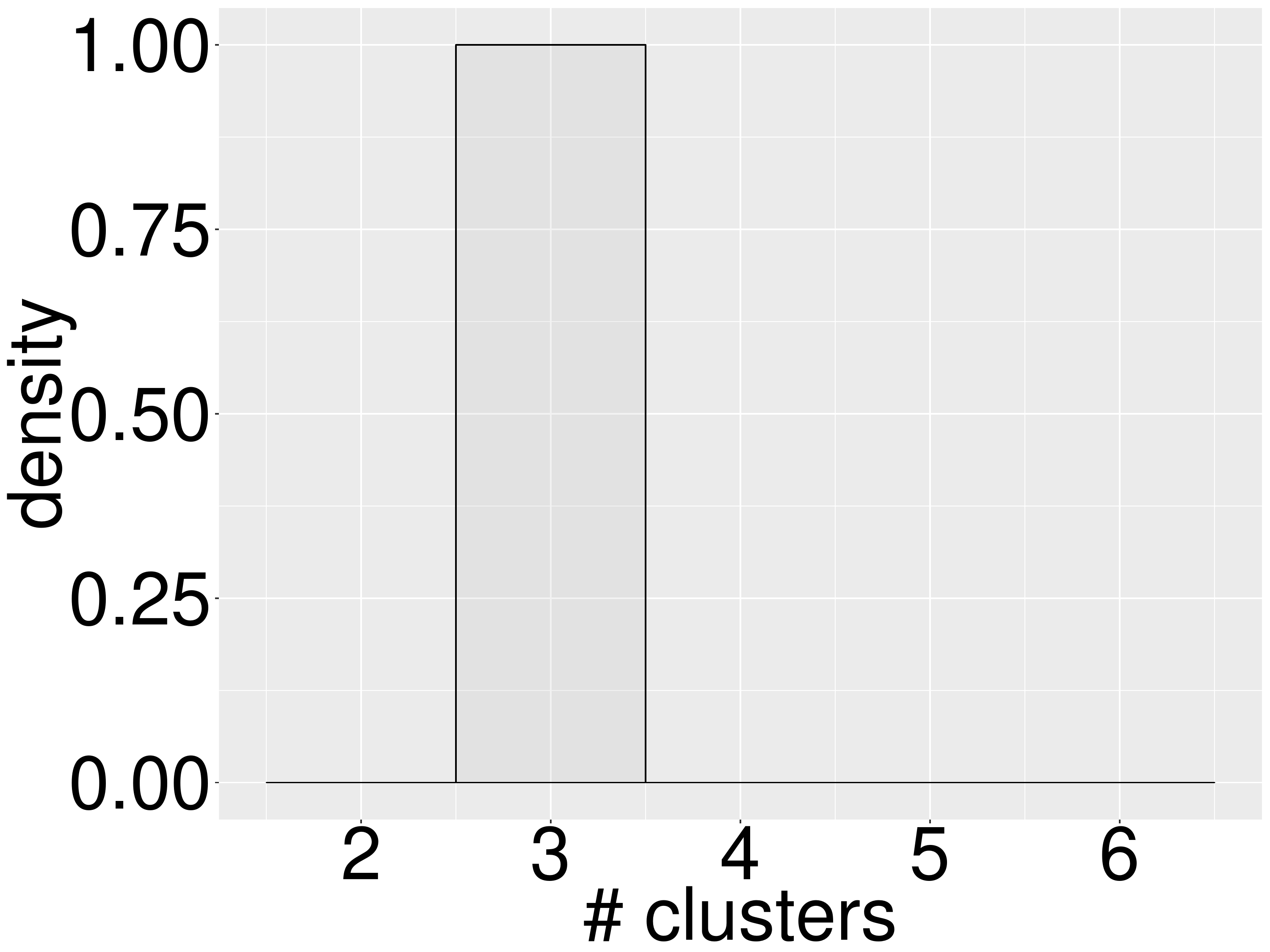} &\includegraphics[width=.33\textwidth]{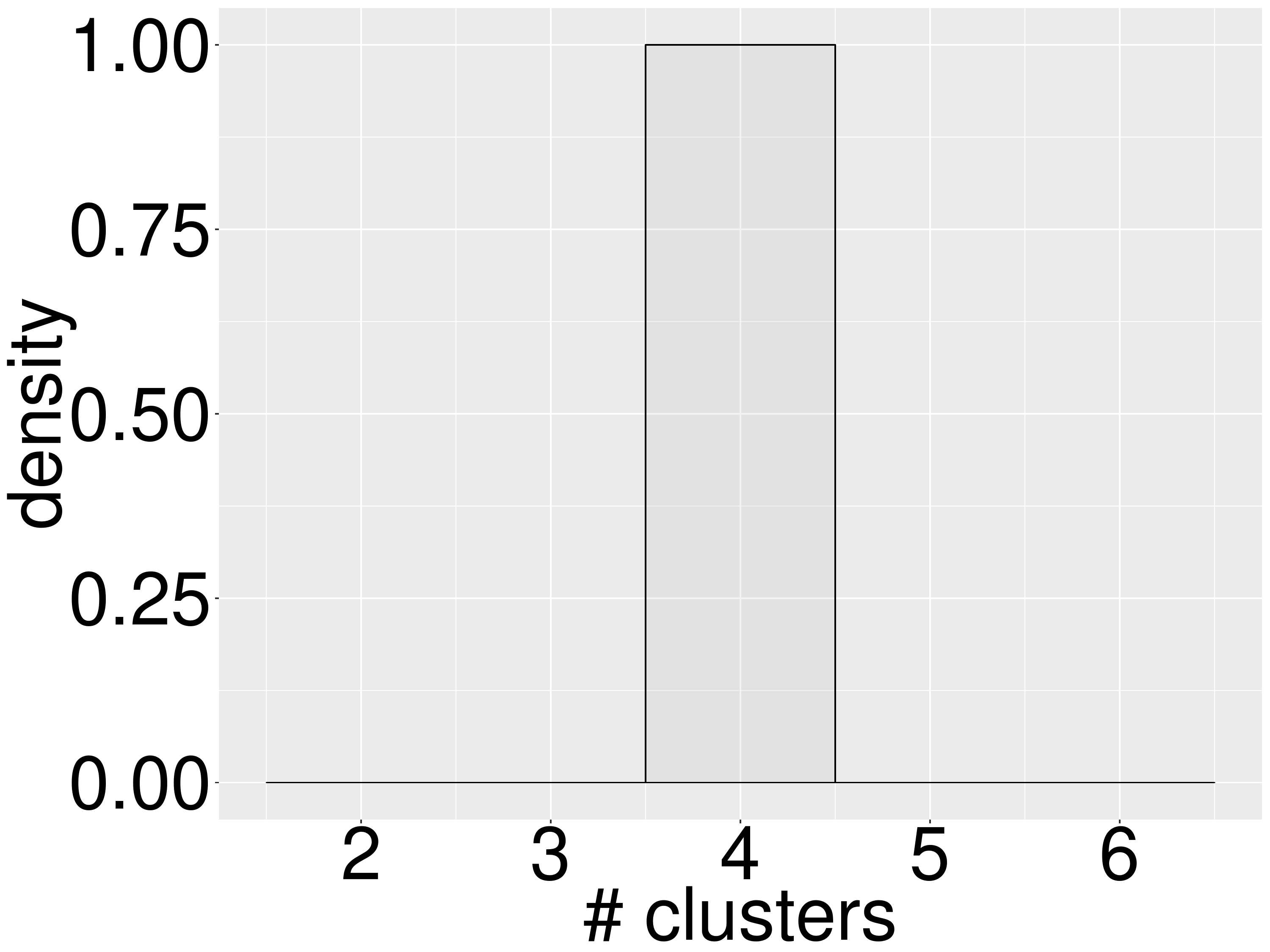}  \\
(a) BAREB $\hat{D}_1$ &(b) BAREB $\hat{D}_2$ & (c) BAREB $\hat{D}_3$ \\
\includegraphics[width=.33\textwidth]{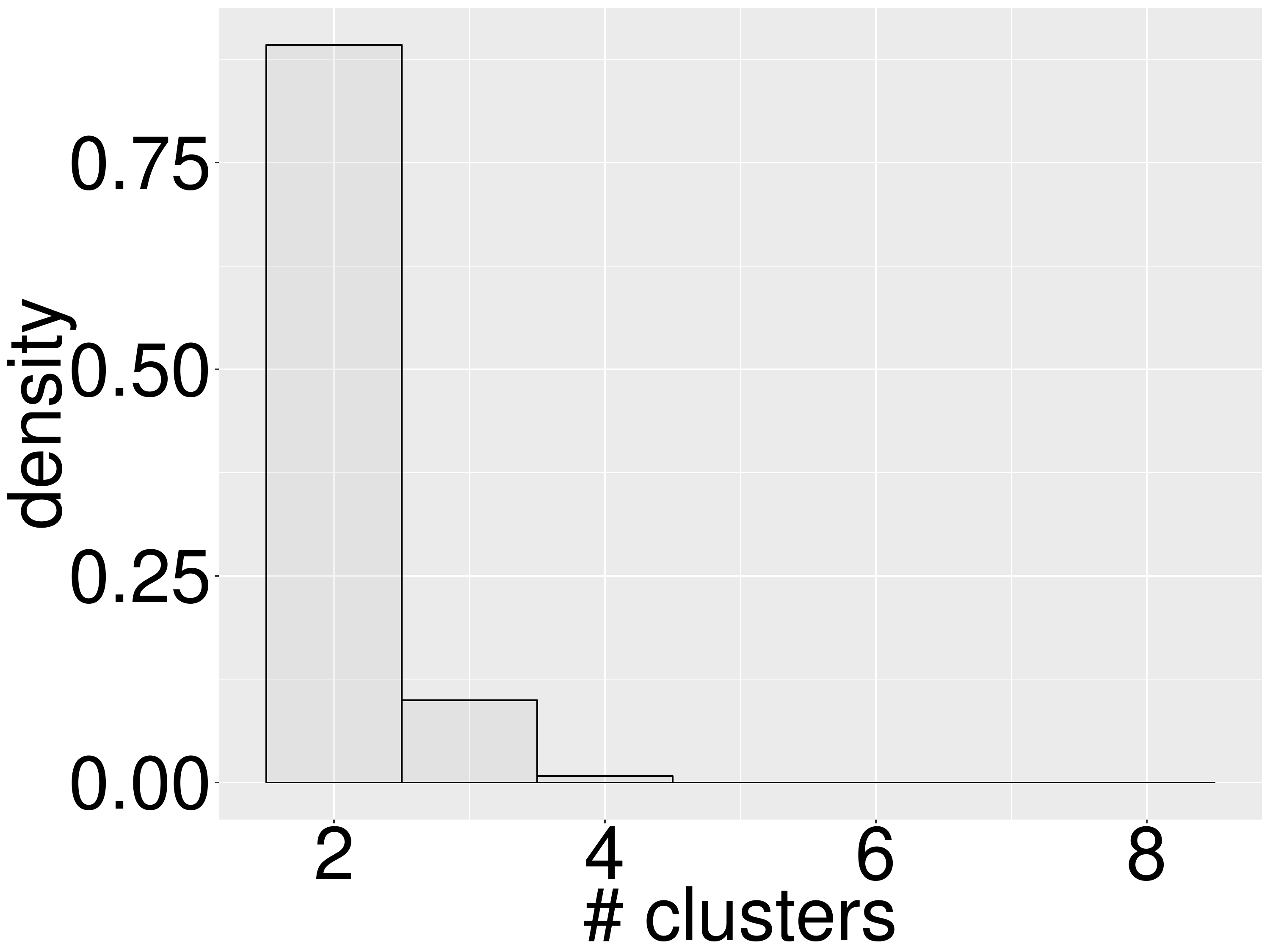} &\includegraphics[width=.33\textwidth]{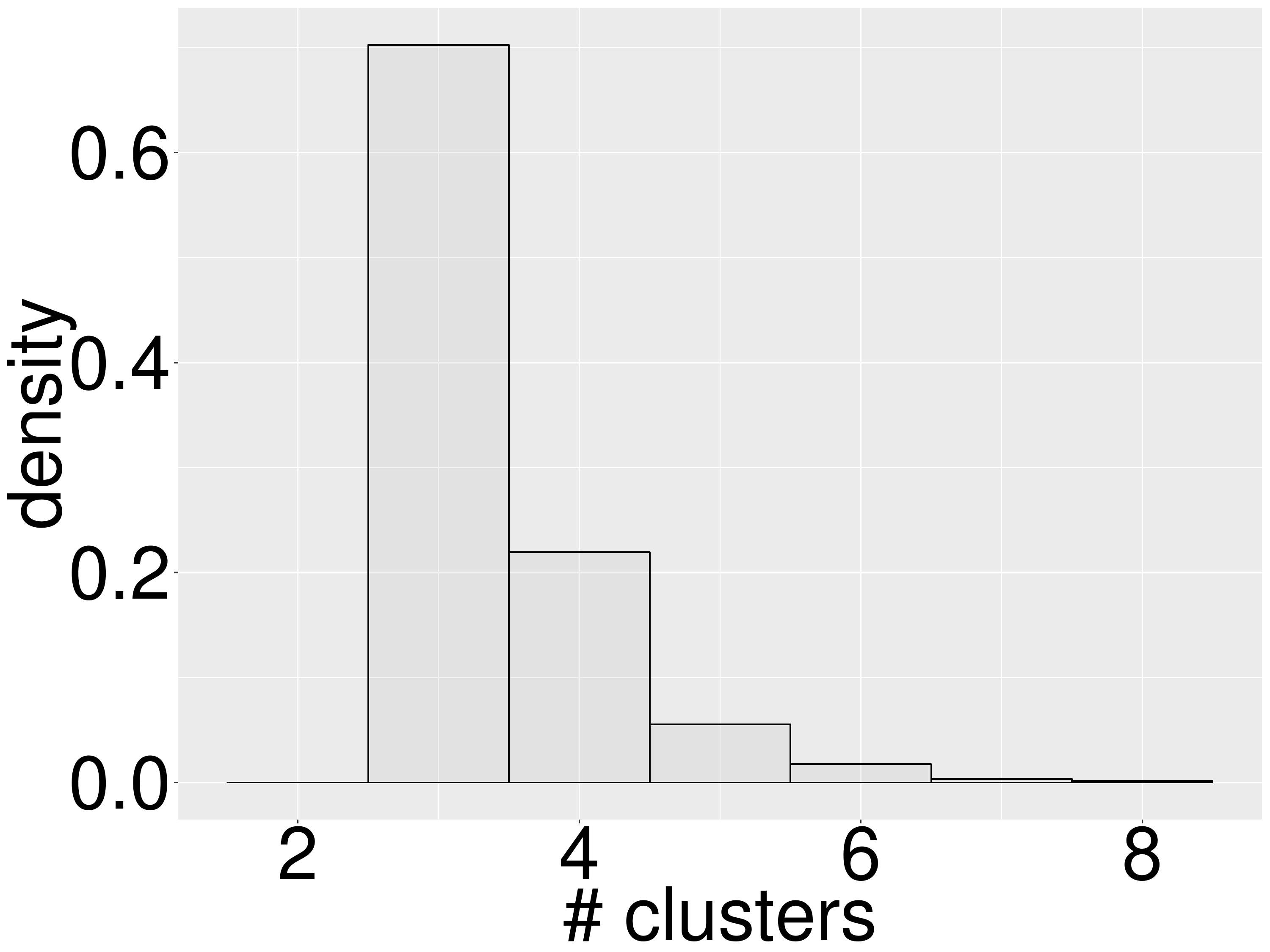} & \includegraphics[width=.33\textwidth]{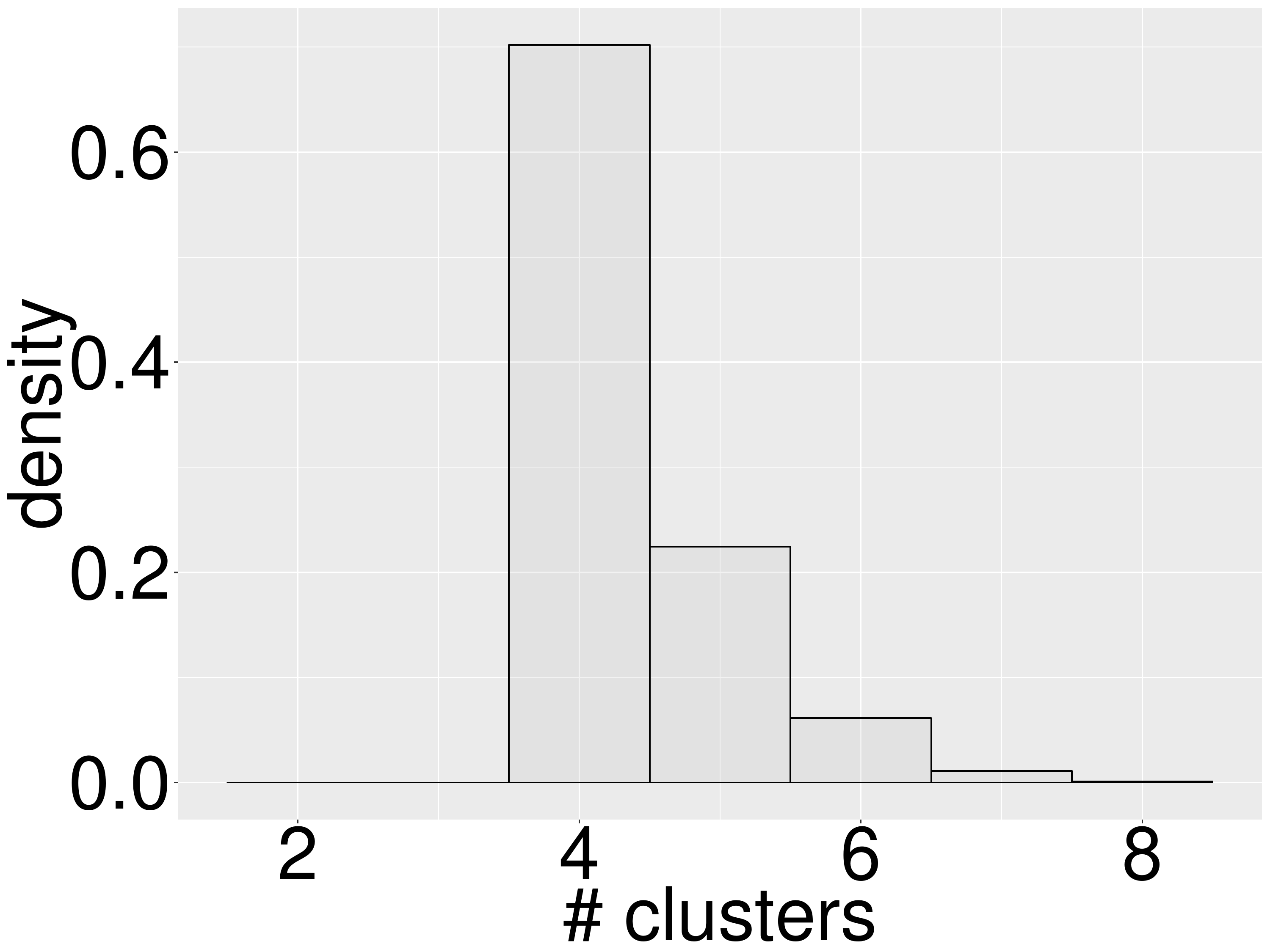}    \\
(d) Indep $\hat{D}_1$ &(e) Indep $\hat{D}_2$ & (f) Indep $\hat{D}_3$ \\
\end{tabular}
\captionsetup{labelformat=empty}
\caption{Figure F3: Comparing site-level clustering from simulated data. Panels(a)-(c) represent histograms of the site cluster cardinality within each subject cluster under \textit{BAREB}, while panels (d-f) represent the same within each subject cluster under the \textit{Indep} model.}
\label{fig:Ks}
\end{figure}

\newpage

\begin{figure}[!tpb]
\begin{centering}
\includegraphics[scale=0.6]{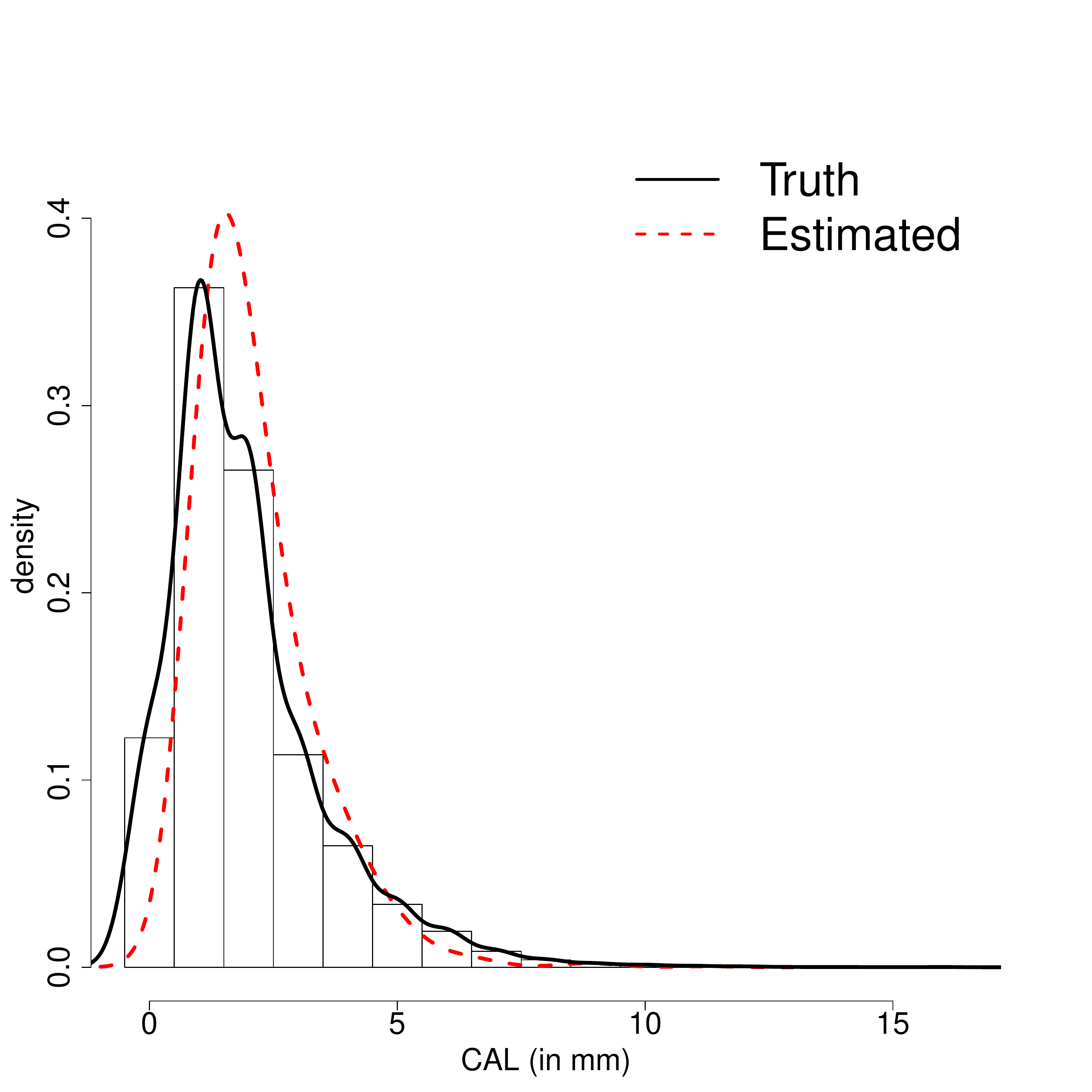}\\
\end{centering}
\captionsetup{labelformat=empty}
\caption{Figure F4: GAAD Data: Density histogram plot of the CAL response, overlaid with the observed and fitted curves generated using Silverman's `rule-of-thumb (ndr0)' smoothing bandwidth.}
\label{fig:predictive_sim}
\end{figure}


\begin{table}[ht]
\centering
\begin{tabular}{ |c|c|c|c| }
\hline
 & Truth & Posterior Mean& MSE\\ \hline
$\widetilde{\bbeta}_1$  & (0.5,1.5,0.5) & (0.488, 1.415, 0.483)  & (3.172e-4, 1.241e-2, 4.000e-4) \\
\hline
$\widetilde{\bbeta}_2$  & (1.5, 2.5, 2.0) & (1.460, 2.297, 1.985) & (1.854e-3, 4.945e-2, 5.178e-4)\\
\hline
$\widetilde{\bbeta}_3$ & (2.5, 3.0, 3.5) & (2.532. 3.067, 3.449) &(1.162e-3, 9.935e-3. 2.782e-3) \\
 \hline
$\widetilde{\bgamma}_{11}$  & (0.0 ,0.5, 0.5) & (0.293, 0.446, 0.531) & (1.052e-1, 5.012e-3, 1.222e-3)  \\
\hline
$\widetilde{\bgamma}_{12}$   & (1.5, 2.0, 2.0) & (1.368, 2.074, 2.019) & (4.972e-2, 9.888e-3, 7.423e-4)\\
\hline
$\widetilde{\bgamma}_{21}$  & (0.0 ,0.5, 0.5) & (0.098, 0.601, 0.525) &(3.187e-2, 1.286e-2, 1.625e-3) \\
\hline
$\widetilde{\bgamma}_{22}$  & (1.5, 2.0, 2.0) & (1.820, 2.051, 2.004) & (2.006e-1, 1.772e-2, 9.921e-4)  \\
\hline
$\widetilde{\bgamma}_{23}$   & (3.0, 4.5, 3.5) & (3.566, 4.415, 3.457) & (3.785e-1, 1.505e-2, 2.489e-3)  \\
\hline
$\widetilde{\bgamma}_{31}$   & (0.0, 0.5, 0.5) & (0.005, 0.470, 0.459) & (2.809e-2, 5.346e-3, 2.445e-3) \\
\hline
$\widetilde{\bgamma}_{32}$   & (1.5, 2.0, 2.0)  & (1.747, 1.901, 2.005) &(1.134e-1, 1.544e-2, 5.115e-4) \\
\hline
$\widetilde{\bgamma}_{33}$   & (3.0, 4.5, 3.5) & (3.177, 4.435, 3.457) & (8.423e-2, 9.874e-3, 2.576e-3)\\
\hline
$\widetilde{\bgamma}_{34}$  & (4.5, 8.0, 5.0) & (4.542, 7.944, 5.009) & (7.234e-2, 1.183e-2, 8.349e-4)  \\
\hline
\end{tabular}
\captionsetup{labelformat=empty}
\caption{Table T1: Simulation Data. First column denotes the simulation truth for subject- and site-level parameters $\{\widetilde{\bbeta}_s\}_{s=1}^{S_0}$ and $\{\widetilde{\bgamma}_{sd}\}_{d=1,s=1}^{D_{s0}, \ \ S_0}$. Columns 2 \& 3 represent the posterior mean and mean-squared error of these parameters estimated under \textit{BAREB}.}
\label{table:betagamma}
\end{table}

\newpage

\begin{table}[!h]
\begin{centering}
\begin{tabular}{p{10cm}r}
 & Patients($n=288$)\\
\hline
\makebox[3cm][c]{Age (years)} & \\
\makebox[3cm][r]{$\leq35$}  & 10 (3\%)\\
\makebox[3cm][r]{ $36-45$}  & 49 (17\%)\\
\makebox[3cm][r]{ $46-55$}  & 82 (29\%)\\
\makebox[3cm][r]{ $56-65$}  & 104 (36\%)\\
\makebox[3cm][r]{ $66-75$}  & 36 (13\%)\\
\makebox[3cm][r]{$\geq76$}  &  7 ( 2\%)\\
\\
\makebox[3cm][c]{Gender} & \\
\makebox[3cm][r]{Male  }  & 69 (24\%)\\
\makebox[3cm][r]{Female}  & 219 (76\%)\\
\\
\makebox[3cm][c]{Smoker} & \\
\makebox[3cm][r]{Yes}  & 88 (31\%)\\
\makebox[3cm][r]{No}  & 200 (69\%)\\
\\
\makebox[3cm][c]{Hb1Ac} & \\
\makebox[3cm][r]{High}  & 170 (59\%)\\
\makebox[3cm][r]{Controlled}  & 118 (41\%)\\
\\
\hline
\end{tabular}
\label{table:pat}
\captionsetup{labelformat=empty}
\caption{Table T2: Subjects characteristics in the motivating GAAD dataset. }
\end{centering}
\end{table}


\begin{table}[ht]
\centering
\begin{tabular}{ |c|c|c| }
\hline
 Parameter & Posterior Mean & 95\% Credible Interval \\ \hline
$c_0$   & -1.211  & (-1.291, -1.131) \\
\hline
$c_1$   & 0.339 & (0.307, 0.373)\\
\hline
$\rho$ & 0.843 &(0.838, 0.865) \\
\hline
$\sigma_{sp}^2$ & 0.026 & (0.025, 0.026)\\
\hline
$\sigma^2$   & 1.545 & (1.517, 1.566)  \\
\hline
\end{tabular}
\captionsetup{labelformat=empty}
\caption{Table T3: Posterior means and 95\% credible intervals for parameters in real data analysis.}
\label{table:real}
\end{table}

\clearpage

\newpage

\section*{\textbf{Web Appendix B: Details on MCMC computing}}

\renewcommand{\theequation}{B-\arabic{equation}}
\setcounter{equation}{0}  


\subsection*{B1. Update cluster indicator prior}
$$\bm{w}\sim \text{Dir}(\alpha_{1} + n_{1},\cdots, \alpha_{S} + n_{S}),$$
and
$$\bm{\phi_{s}} \sim \text{Dir}(\alpha_{s1}^{\phi} + n_{s1},\cdots, \alpha_{sD_{s}}^{\phi}+n_{sD_{s}}),$$
where $n_{s}$ is the \# os subjects in the subject cluster $s$; $n_{sd}$ is the \# of sites in site cluster $d$ in the $s$-th subject cluster; $\alpha_{s} = \alpha_{sd}^{\phi}= a  = 1$ for all $s=1,\dots, S$ and $d=1, \dots, D_{s}$.

\subsection*{B2. Update $\bm{e}$, the subject cluster membership indicator}
$$P(e_{i}=s\mid\dots)\propto w_{s}\times \prod_{j=1}^{J}[f(y_{ij}\mid e_{i}=s,...)\times\prod_{t=1}^{T}P(\delta_{i}(t)\mid e_{i}=s,\dots)],$$
where
$$f(y_{ij}|e_{i}=s,\ldots )=N(y_{ij}; \mu_{ij},\sigma^{2}),$$
$$P(\delta_{i}(t)=1\mid e_{i}=s,\ldots)=\Phi(\mu_{i}^{*}(t)),$$
$$\mu_{ij}=\bm{x}_{i}\widetilde{\bbeta}_{s}+\bm{z}_{j}\widetilde{\bgamma}_{sr_{sj}} + \nu_{ij},$$
$$\mu_{i}^{*}(t)= c_0 + c_1 R_{t}\bm{\mu}_{i}.$$

\subsection*{B3. Update $\bm{r}$, the vector of clustering labels}
Consider the site-level clustering for subject cluster $s$. Let $A_{s}$ be the set of subjects assigned to $s$-th subject cluster, then

$$P(r_{sj}=r\mid\dots)\propto \alpha_{sr}^{\phi}\times \prod_{i\in A_{s}}[f(y_{ij}|r_{sj}=r,...)\times P(\delta_{i}(t_{j})|r_{sj}=r,...)],$$
where $t_{j}$ denotes site $j$ on tooth $t$,

$$f(y_{ij}|r_{sj}=r,\dots)=N(y_{ij}; \mu_{ij},\sigma^{2}),$$
$$P(\delta_{i}(t_j)=1\mid r_{sj}=r,\dots)=\Phi(\mu_{i}^{*}(t)),$$
$$\mu_{ij}=\bm{x}_{i}\widetilde{\bbeta}_{s}+\bm{z}_{j}\widetilde{\bgamma}_{sr} + \nu_{ij},$$
$$\mu_{i}^{*}(t)= c_0 + c_1 R_{t}\bm{\mu}_{i}.$$

\subsection*{B4. Update parameters $\{\widetilde{\bbeta}_s\}_{s=1}^S$ and $\{\widetilde{\bgamma}_{sd}\}_{d=1,s=1}^{D_s, \ \ S}$}
Unfortunately, we do not have conjugate priors for $\{\widetilde{\bbeta}_s\}_{s=1}^S$, hence, we resort to Metropolis-Hastings updates. We update one atom of $\{\widetilde{\bbeta}_s\}_{s=1}^S$ each time. In practice, we might update entries of $\widetilde{\bbeta}_s$ iteratively to increase the acceptance ratio. We propose a new $\widetilde{\bbeta}'_s$ from distribution
$$\widetilde{\bbeta}'_s\mid \widetilde{\bbeta}_s \sim N(\widetilde{\bbeta}_s,\sigma_{\beta}^{2}I).$$
Then, the acceptance ratio is $\text{min}(1,\alpha_{\beta})$, where
$$\alpha_{\beta}=\frac
{\pi(\{\widetilde{\bbeta}_1,\dots, \widetilde{\bbeta}_{s-1},\widetilde{\bbeta}'_s,\widetilde{\bbeta}_{s+1},\dots, \widetilde{\bbeta}_{S}\})f(Y\mid \widetilde{\bbeta}'_s)P(\Delta\mid \widetilde{\bbeta}'_s)}
{\pi(\{\widetilde{\bbeta}_s\}_{s=1}^S)f(Y\mid \widetilde{\bbeta}_s)P(\Delta\mid \widetilde{\bbeta}_s)},$$
where $\widetilde{\bbeta}'_s$ is the new set of unique values with $\bm{\beta}_{s}^{*}$ replaced by $\bm{\beta}_{s}^{'*}$. Note, although $f(Y\mid\bm{\beta}^{'*})\text{ and }f(Y\mid\bm{\beta}^{*})$ are the likelihoods of all observed CAL values, most of their contributions cancels out, with the remaining likelihood only involving subjects in cluster $s$. Similar is true for $P(\Delta\mid\bm{\beta}^{'*})$ and $P(\Delta\mid\bm{\beta}^{*})$.

Consider the prior ratio $\pi(\bm{\beta}^{'*})/\pi(\bm{\beta}^{*})$. Let $C_{\beta}^{-s}$ be the submatrix of $C_{\beta}$ with the $s$th row and $s$th column removed; and $c_{\beta}^{-s}$ be the $s$th row of $C_{\beta}$ with $s$th entry removed. Then, we would have
$$\pi(\bm{\beta}^{'*})/\pi(\bm{\beta}^{*}) =  \frac
{C(\beta_{s}^{'*},\beta_{s}^{'*})-c_{\beta}^{-s}(C_{\beta}^{-s})^{-1}(c_{\beta}^{-s})^{T}}
{C(\beta_{s}^{*},\beta_{s}^{*})-c_{\beta}^{-s}(C_{\beta}^{-s})^{-1}(c_{\beta}^{-s})^{T}}.
$$

Updating $\bm{\gamma}$ is similar to updating $\bm{\beta}$, with the only exception that while updating $\bm{\gamma}_{sj}^{*}$, the likelihood only involves subjects in cluster $s$ and sites in cluster $j$.

\subsection*{B5. Update $\bm c$, the parameter that controls the relationship between CAL, and the non-random missingness indicator}
Assume $N(\bm 0,\sigma_{c}^{2}\bm I)$ as the prior for $\bm c = (c_0, c_1)$, then using standard linear regression with white noise, we have
\[
\bm c\mid \cdots
\sim \mathcal{N}(
\begin{bmatrix}
    \sum_{i=1}^{N}\sum_{t=1}^{T} g_i(t) \\
    \sum_{i=1}^{N}\sum_{t=1}^{T} (R_{t}\bm{\mu}_{i})g_i(t)
\end{bmatrix},
\begin{bmatrix}
    NT  & \sum_{i=1}^{N}\sum_{t=1}^{T} (R_t \bm{\mu}_{i})\\
    \sum_{i=1}^{N}\sum_{t=1}^{T} (R_t \bm{\mu}_{i})& \sum_{i=1}^{N}\sum_{t=1}^{T}(R_t \bm{\mu}_{i})^2
\end{bmatrix}
)
\]

\subsection*{B6. Update $\mu$ and $\mu^{*}$}
It simply requires computing $\mu$ and $\mu^{*}$ with updated parameters:
$$\mu_{ij}=\bx_{i}\bbeta_{i}+\bz_{j}\bgamma_{ij} + \nu_{ij},$$
$$\mu_{i}^{*}(t) =  c_0 + c_1 R_{t}\bm{\mu}_{i}$$

\subsection*{B7. Update $\nu_{ij}$ and its hyperparameter}
Let $\bm\nu_{i} = (\nu_{i1},\dots, \nu_{iJ})^T$, for $i=1,\dots, N$ and $\bm\nu = (\bm\nu_1,\dots, \bm\nu_N)^T$. We assume
$$\bm\nu_{1},\dots, \bm\nu_N\iid \text{MVN}(0,\bm\Sigma),$$
where  $\bm\Sigma$ is assigned a conditional autoregressive (CAR) prior: $\bm\Sigma = \sigma_{sp}^2 \bm G(\rho)^{-1}$, where $\bm G(\rho) = \bm B-\rho \bm W$. Here $\bm B$ is a diagonal matrix with $j$th diagonal entry being the number of neighbors at $j$th site and $\bm W$ denotes the adjacency matrix for 168 sites in the mouth structure. 

Assuming $\sigma_{sp}^2\sim \text{IG}(a_{sp}, b_{sp})$, we have
$$f(\sigma_{sp}^2\mid \dots)\propto (\sigma_{sp}^2) ^{-a_{sp}-1}\text{exp}\{-\frac{b_{sp}}{\sigma_{sp}^2}\}
\prod_{i=1}^N \text{det}(\sigma_{sp}^2 \bm G(\rho)^{-1})^{-1/2} \text{exp}\{-\frac{1}{2}\frac{\bm\nu_i^T\bm G(\rho)\bm\nu_i}{\sigma_{sp}^2}\}. $$
Therefore,
$$f(\sigma_{sp}^2\mid \dots)\propto (\sigma_{sp}^2) ^{-a_{sp} - \frac{NJ}{2}-1} \text{exp}\{-\frac{1}{\sigma_{sp}^2}(b_{sp} + \sum_i \frac{\nu_i^T\bm G(\rho)\bm\nu_i}{2})\}.$$
$$\sigma_{sp}^2\mid \dots\sim \text{IG}(a_{sp}+\frac{NJ}{2}, b_{sp} + \sum_i\frac{\nu_i^T\bm G(\rho)\bm\nu_i}{2} ).$$

We assign a uniform prior on $\rho$: $\rho\sim \text{Unif}(a_{\rho}, b_{\rho})$ and update $\rho$ using a
 Metropolis-Hastings sampler. With a
  step size $\sigma_{\rho}$, we propose a new $\rho$ from $\rho'\mid\rho\sim \text{Unif}\big [\text{max}(a_{\rho}, \rho+\sigma_\rho), \text{min}(b_{\rho},  \rho-\sigma_\rho)\big ]$.
The acceptance ratio is $\text{min}(1,\alpha_\rho)$, where
$$\alpha_\rho = \frac{f(\rho\mid\rho')}{f(\rho'\mid\rho)} \frac{f(\rho'\mid\dots)}{f(\rho\mid\dots)}.$$
Here,
$$ \frac{f(\rho\mid\rho')}{f(\rho'\mid\rho)} = \frac{\text{min}(\rho+\sigma_\rho, b_{\rho}) - \text{max}(\rho-\sigma_\rho, a_{\rho})}{\text{min}(\rho'+\sigma_\rho, b_{\rho}) - \text{max}(\rho'-\sigma_\rho, a_{\rho})},$$
$$f(\rho\mid \dots)\propto \mathbb{I}_{(a_{\rho}, b_{\rho})} \prod_{i=1}^N \text{det}(\bm G(\rho)^{-1})^{-1/2} \text{exp}\{-\frac{1}{2}\frac{\bm\nu_i^T\bm G(\rho)\bm\nu_i}{\sigma_{sp}^2}\}. $$

To update each entry of $\bm\nu$, $\nu_{ij}$, we use Metropolis-Hastings algorithm and with the proposal distribution: $\nu_{ij}'\mid\nu_{ij}\sim N(\nu_{ij}, \sigma_\nu^2)$.
The acceptance ratio is $\text{min}(1,\alpha_\nu)$, where
$$\alpha_\nu = \frac{\pi(\nu_{i1},\dots, \nu_{i,j-1},\nu_{ij}',\nu_{i,j+1},\dots, \nu_{iJ}\mid\bm\Omega)f(y_{ij}\mid \nu_{ij}')P(\delta_i(t_j)\mid \nu_{ij}')}{\pi(\bm\nu_i\mid\bm\Omega)f(y_{ij}\mid \nu_{ij})P(\delta_i(t_j)\mid \nu_{ij})}.$$
Recall that
$$f(y_{ij})=N(y_{ij}; \mu_{ij},\sigma^{2}),$$
$$P(\delta_{i}(t_j)=1)=\Phi(\mu_{i}^{*}(t)),$$
$$\mu_{ij}=\bm{x}_{i}\bbeta_{j}+\bm{z}_{j}\bgamma_{ij} + \nu_{ij},$$
$$\mu_{i}^{*}(t)= c_0 + c_1 R_{t}\bm{\mu}_{i}.$$

\subsection*{B8. Update $g_i(t)$}
We generate $g_i (t)$ from truncated normal distribution. To be more specific, if tooth $t$ of subject $i$ is missing, then
$$g_i (t)\sim N(\mu_{it}^{*},1)I_{\{g_i (t)>0\}},$$
otherwise,
$$g_i (t)\sim N(\mu_{it}^{*},1)I_{\{g_i (t)<0\}}.$$

\subsection*{B9. Update $\theta_\beta$ and $\theta_{\gamma_s}$}
We assume a non-informative prior for both $\theta_\beta$ and $\theta_{\gamma_s}$: $N(0,100)$. To update $\theta_\beta$, we use Metropolis-Hastings algorithm with the proposal distribution $\theta_\beta' \mid \theta_\beta \sim N(\theta_\beta, 0.1^2)$.
The acceptance ratio is $\text{min}(1,\alpha_{\theta_\beta})$, where
$$\alpha_{\theta_\beta} = \frac{p(\{\widetilde{\bm\beta}_s\}_{s=1}^S\mid\theta_\beta',\ S) p(\theta_\beta')}
{p(\{\widetilde{\bm\beta}_s\}_{s=1}^S\mid\theta_\beta,\ S) p(\theta_\beta)}.$$
Recall equation (3.1) for computing the likelihood of $\{\widetilde{\bm\beta}_s\}_{s=1}^S$.

The $\theta_{\gamma_s}$ is updated in the same way.

\subsection*{B10. Update $D_s$}
We use RJMCMC \citep{green1995reversible} to update the \# of tooth-site level clusters $D_{s}$. We propose to either combine two clusters into one, or split one cluster into two. Note, during matching the mean $\mu_{ij}=\bx_{i}\bbeta_{i}+\bz_{j}\bgamma_{ij} + \nu_{ij}$, the $\bx_{i}\bbeta_{i}$ and $\nu_{ij}$ term cancels out, and only $\bz_{j}\bgamma_{ij}$ term is of interest.
First, we choose the move type. We propose a merge move with probability $q_{Dd}$ and a split move with probability $q_{Du}$, where $q_{Du} = 1- q_{Dd}$, and

\[
   q_{Dd}=
\begin{cases}
     1, & \text{if } D_s=D_{max}=10\\
     0.5 ,& \text{if } D_{max}>D_s\geq 2\\
      0,         & \text{if } D_s=1
\end{cases}
\]

If a combine move is chosen, we randomly select $d_{1},\ d_{2}$ from $\{1,\ldots, D_s\}$ with probability $\frac{1}{D_s(D_s-1)}$. Without loss of generality, suppose $d_{1}=1$ and $d_{2}=2$. With little abuse of notations, we let $w_{1}$ and $w_{2}$ be the probability of one site being in cluster 1 and 2, respectively. Denote $\tilde{w}_1$, the probability of one site being in the new combined cluster, and $\tilde{\bm{\gamma}}_1$, the corresponding linear coefficient. Then, for each site $i$, we want
$$w_{1}+w_{2} = \tilde{w}_{1},$$
$$w_{1}\bm{z}_{i}\bm{\gamma}_{1}+w_{2}\bm{z}_{i}\bm{\gamma}_{2} = \tilde{w}_{1}\bm{z}_{i}\tilde{\bm{\gamma}}_{1}.$$
However, it suffices to have
$$w_{1}+w_{2} = \tilde{w}_{1},$$
$$w_{1}\bm{\gamma}_{1}+w_{2}\bm{\gamma}_{2} = \tilde{w}_{1}\tilde{\bm{\gamma}}_{1}.$$

\noindent Thus, we have
$$\tilde{\bm{\gamma}}_{1} = (w_{1}\bm{\gamma}_{1}+w_{2}\bm{\gamma}_{2})/\tilde{w}_{1}$$

If a split move is chosen, we randomly select $d$ from $\{1,\ldots, D_s\}$ with probability $\frac{1}{D_s}$. Without loss of generality, suppose $d=1$. Suppose, after split, a site can be in one of the new clusters with probabilities $\tilde{w}_1$ and $\tilde{w}_2$. Denote $\tilde{\bm{\gamma}}_1$ and $\tilde{\bm{\gamma}}_2$,  the corresponding linear coefficients for the new clusters. They are computed as follows:

$$\tilde{w}_{1} = w_{1}\alpha\ \ \ \ \ \tilde{w}_{2} = w_{1}(1-\alpha)\text{, where }\alpha\sim Beta(1,1).$$

\noindent The $k$th element in $\bm{\gamma}$:
$$
\tilde{\gamma}_{1k}=\gamma_{1k}-\sqrt{\frac{\tilde{w}_{2}}{\tilde{w}_{1}}}\beta_{k}
=\gamma_{1k}-\sqrt{\frac{1-\alpha}{\alpha}}\beta_{k}, $$
$$
\tilde{\gamma}_{2k}=\gamma_{1k}+\sqrt{\frac{\tilde{w}_{1}}{\tilde{w}_{2}}}\beta_{k}
=\gamma_{1k}+\sqrt{\frac{\alpha}{1-\alpha}}\beta_{k}, $$
where $\beta_{k}\overset{i.i.d}{\sim} Beta(2,2).$

\noindent Consider the Jacobian $J$: $J = \bigg(\frac{\partial \bm{\theta}^{(2)}}{\partial(\bm{\theta}^{(1)},\bm{u}^{(1)})} \bigg)$, where $\bm{\theta}^{(2)} = (\tilde{w}_{1},\tilde{w}_{2},\tilde{\bm{\gamma}}_{1},\tilde{\bm{\gamma}}_{2})$, $\bm{\theta}^{(1)} = (w_{1},\bm{\gamma}_{1})$, $\bm{u}^{(1)} = (\alpha, \bm{\beta})$, with $\bm{\gamma}_{1},\ \tilde{\bm{\gamma}}_{1},\ \tilde{\bm{\gamma}}_{2},\ \bm{\beta}\in \mathbb{R}^{l}$. Here $l$ is the number of site-level parameters, and

\[
J
=
\begin{bmatrix}
    \alpha & w_{1} & 0 & 0 \\
    1-\alpha & -w_{1} & 0 & 0 \\
    \text{*} & \text{*} & I & -\sqrt{\frac{1-\alpha}{\alpha}} I \\
    \text{*} & \text{*} & I & \sqrt{\frac{\alpha}{1-\alpha}}I \\

\end{bmatrix},
\] and

\[
|det(J)|
=
\begin{bmatrix}
   \alpha & w_{1} \\
   1-\alpha & -w_{1}
\end{bmatrix}
\cdot
\begin{bmatrix}
    I & -\sqrt{\frac{1-\alpha}{\alpha}} I \\
    I & \sqrt{\frac{\alpha}{1-\alpha}}I \\
\end{bmatrix}
=w_{1}[\alpha(1-\alpha)]^{-\frac{l}{2}}.
\]

\noindent Part of the first two columns of $J$ is represented by $\text{*}$ since it will not affect the determinant computation.
Now, the acceptance ratio of split is:
$$\underbrace{\frac{p(D_s+1,\tilde{w}, \tilde{\bm{\gamma}}\mid Y)}{p(D_s, w,\bm{\gamma}\mid Y)}}_{(1)}
\times
\underbrace{
\frac{q_{D_s+1,d}}{q_{D_s,u}}
\times
\frac{q_{D_s+1,c}(d_{1},d_{2})}{q_{D_s,s}(d)}
\times
\frac{1}{p_{\bm{u}}(\alpha,\bm{\beta})}
}_{(2)}
\times
\frac{|det(J)|}{D_s+1}
$$

\noindent Here, $$(1)=(\text{likelihood ratio}) \frac{p(D_s+1)}{p(D_s)} \frac{p(\tilde{w})}{p(w)}\frac{p(\tilde{\bm{\gamma}})}{p(\bm{\gamma})}, $$
where likelihood ratio depends on both $Y$ and missing indicators $\Delta$, and
$$\frac{p(D_s+1)}{p(D_s)} = \frac{1}{D_s+1}, \ 
\frac{p(\tilde{\bm{\gamma}})}{p(\bm{\gamma})} = \frac{det(C_{D_s+1})}{det(C_{D_s})}, \ 
\frac{p(\tilde{w})}{p(w)} = \frac{\tilde{w}_{1}^{a -1 + \tilde{n}_{1}}\tilde{w}_{2}^{a -1 + \tilde{n}_{2}}}{w_{1}^{a - 1 +\tilde{n}_{1}+ \tilde{n}_{2}} B(a,aD_s )}, \ \mbox{and}$$ \\

$$(2)=
\begin{cases}
     \frac{2}{D_s+1}\frac{1}{p_{\bm{u}}(\alpha,\bm{\beta})}, & \text{if } D_s=D_{max}=10\\
     \frac{1}{D_s+1}\frac{1}{p_{\bm{u}}(\alpha,\bm{\beta})} ,& \text{if } D_{max}>D_s>1\\
      \frac{1}{2(D_s+1)}\frac{1}{p_{\bm{u}}(\alpha,\bm{\beta})},         & \text{if } D_s=1,
\end{cases}$$ \\

\noindent where $det(C_{D_s+1})$ and $det(C_{D_s})$ are the DPP likelihood, $B(\cdot,\cdot)$ is the Beta function, $\tilde{n}_1$ and $\tilde{n}_2$ are the numbers of sites in the new clusters, and $p_{u}(\alpha,\bm{\beta}) = p(\alpha)p(\bm{\beta})p_{0}$, with $p_{0}$ being the probability of relocating sites in the chosen cluster into the two new clusters. The probability is the production of the prior, $\tilde{w}_1$ or $\tilde{w}_2$, and the likelihood, which again depends on both $Y$ and missing indicators $\Delta$. The acceptance ratio of merge is just the inverse of the corresponding split acceptance ratio.


\end{document}